**LETTER of INTENT to PAC**

**CDF/PHYS/EXOTIC/CDFR/5585**

**March 26, 2001**

# A Search for the Higgs Boson using Very Forward Tracking Detectors with CDF.

M.G.Albrow†,[1] M.Atac,[1] P.Booth,[2] P.Crosby,[3] I.Dunietz,[1] D.A.Finley,[1] B.Heinemann,[2] M.Lancaster,[3] R.Lauhakangas,[5] D.Litvintsev,[1] T.Liu,[1] S.Marti-Garcia,[2] D.McGivern,[3] C.D.Moore,[1] R.Orava,[4,5] A.Rostovtsev,[6] R.Snihur,[3] S.Tapprogge,[5] W.Wester,[1] A.Wyatt,[3] K.Österberg.[4]

[1] *Fermi National Accelerator Laboratory, Batavia, IL 60510*

[2] *Liverpool University, England*

[3] *Univ. College London, England*

[4] *Univ. Helsinki, Finland*

[5] *Helsinki Institute of Physics, Finland*

[6] *Inst. Theor. Exp. Phys. (ITEP), Moscow, Russia*

† Contact person

// Abstract

We propose to add high precision track detectors 55 m downstream on both (E & W) sides of CDF, to measure high Feynman-$x$ protons and antiprotons in association with central states. A primary motivation is to search for the Higgs boson, and if it is seen to measure its mass precisely. The track detectors will be silicon strip telescopes backed up by high resolution time-of-flight counters. We will have four spectrometer arms, for both sides of the $p$ and $\bar{p}$ beams. The addition of these small detectors effectively converts the Tevatron into a gluon-gluon collider with $\sqrt{s}$ from 0 to $\approx$ 200 GeV.

This experiment will also measure millions/year clean high- $|t|$ elastic $p\bar{p}$ scattering events and produce millions of pure gluon jets. Besides a wealth of other unique QCD studies we will search for signs of exotic physics such as SUSY and Large Extra Dimensions.

We ask the Director to ask the PAC to take note of this Letter of Intent at its April meeting, to consider a proposal at the June meeting and to make a decision at the November 2001 meeting. We request that the Directorate ask the Beams Division to evaluate the consequences and cost of the proposed Tevatron modifications, and CDF to evaluate any effect on its baseline program and to review the technical aspects of the detectors, DAQ and trigger integration.

## Contents

## I. INTRODUCTION

Two prominent areas of uncertainty in the Standard Model are the Higgs sector and non-perturbative QCD with the problem of confinement. Both of these are intimately related to properties of the vacuum, either electroweak or strong, and both are subjects of this letter of intent.

As the Higgs boson $H$ has vacuum quantum numbers it can be produced in the exclusive reaction $p\bar{p} \to pH\bar{p}$. In the most probable symmetric situation the $p$ and $\bar{p}$ each lose about $\frac{M_H}{2}$ of energy and have small $p_T$, and the $H$ has low rapidity and decays (isotropically of course) into the central CDF detector. The process is the dominant $gg \to H$ process through a top quark loop (we expect 10,500 such $H$ in 15 fb$^{-1}$ for $M_H = 120$ GeV) together with soft gluon exchanges that cancel the color removed from the $p$ and $\bar{p}$ and can even leave them in their ground state. We know the incoming beam momenta and will measure the outgoing $p$ and $\bar{p}$ momenta very precisely ($\frac{\delta p}{p} \approx 10^{-4}$) and so can reconstruct $M_H$ as the *Missing Mass* $MM$ to the $p$ and $\bar{p}$

$$MM^2 = (p_{b1} + p_{b2} - p_3 - p_4)^2$$

where the $p$ are 4-vectors with obvious notation. This will be done when the central system looks like $b\bar{b}$, $\tau^+\tau^-$ or $W^+W^-$. Clearly missing neutrinos and poor central mass resolution do not affect the missing mass measurement, for which we expect $\sigma_{MM} \approx 250$ MeV independent of $MM$. Such good resolution reduces continuum background and means that if a signal is seen we can measure $M_H$ extremely well (e.g. to 80 MeV (statistical) for 10 events). There is much uncertainty on the value for this exclusive cross section $\sigma(p\bar{p} \to pH\bar{p})$ because it involves non-perturbative QCD. Predictions range from more than 100 events (on very little background) for a 120 GeV Higgs in 15 fb$^{-1}$ to about 1 event. Part of the controversy is whether associated hadrons (perhaps central) will accompany the Higgs. For the $\tau^+\tau^-$ and $WW^{(*)} \to leptons$ modes any associated hadrons, if well measured, can be subtracted out in the misssing mass sum

$$MM^2 = (p_{b1} + p_{b2} - p_3 - p_4 - \sum_{i=5}^{n} p_i)^2.$$

This can not of course be done with the $H \to b\bar{b}$ decay because we cannot distinguish the associated hadrons from the $H$-decay products. Also we are ignoring smearing due to particles not detected, including soft photons (QED bremsstrahlung).

There is a normalizing process which is very similar to the exclusive Higgs production process, namely $p\bar{p} \to p\gamma\gamma\bar{p}$. Both proceed through $gg \to$ quark loops; in the first case it is mostly top and in the second mostly up-quark. Neither the Higgs nor the $\gamma\gamma$ couple to first order through the strong interaction to the rest of the event, and a 120 GeV Higgs is so long-lived ($\Gamma_H$ <3 MeV) that its decay products cannot affect the primary interaction. (If the Higgs is heavy we can use $WW \to$ leptons.) We cannot measure $\gamma\gamma$ near $M_H$ but will be able to look for exclusive $p\bar{p} \to p\gamma\gamma\bar{p}$ in the mass range 10 - 40 GeV. The different $Q^2$ is an issue which may have to be taken into account using theory. Central dijet production with a leading $p$ and $\bar{p}$ is *not* a good normalizer for the exclusive process because *all* the particles involved are strongly interacting.

In addition to the Higgs search, beyond Standard Model physics (light gluino $\tilde{\chi}_1^\circ$ pairs if there is SUSY, gravitons or Micro Black Holes if there are Large Extra Dimensions, etc.) is potentially accessible by this missing mass method. While we will look for such exotica, we do not expect that the Tevatron has enough energy to see signals. Nevertheless we will be pioneering the technique, which may be successful at the LHC or VLHC.

While the search for Higgs is our primary motivation, there are many unresolved questions of the strong interaction which we will address. Confinement is to do with how quarks and gluons end up, in *every collision* involving hadrons, in color singlet "clumps" i.e. hadrons. In the transition sometimes very massive color singlet clumps are formed, well separated from each other in rapidity space. The physics of rapidity gaps is closely related to diffraction; the largest gaps at the Tevatron being 15 units in elastic scattering.

We have made many hard (high $Q^2$) studies in CDF of diffraction and rapidity gaps, finding diffractively produced jets [1], $b$-jets [2], $W$ [3], and $J/\psi$ [4]. In Run 1C we used roman pots with scintillating fibre hodoscopes to measure high-$x_F$ antiprotons and used jets to measure the diffractive structure function of the $\bar{p}$ [5], and we discovered double pomeron exchange ($DI\!PE$) production of high-$E_T$ jets [6].

For Run 2A we are re-instrumenting (with new electronics) the previously used roman pots (which are only on the $\bar{p}$ side), we have installed new Beam Shower Counters (BSC) to tag forward rapidity gaps ($5.5 < \eta < 7.0$), and will install Miniplug calorimeters with high transverse granularity in the rapidity region $3.5 < |\eta| < 5.5$. The physics program is summarized in the proposal for experiment E916 [7].

Although not our primary motivation, there is much unique strong interaction physics that can be done with this proposed addition to CDF. Some of this is outlined in Appendix I.

This is a letter of intent to supplement CDF with very forward tracking detectors to measure both $p$ and $\bar{p}$ in events where they have fractional momentum loss $\xi = 1.0 - \frac{p_{out}}{p_{in}}$ in the range up to approximately 0.10. Knowing the beam 4-momenta, $p_{b1}$ and $p_{b2}$, and the outgoing 4-momenta $p_3$ and $p_4$, we calculate the missing mass $MM$.

## II. APPARATUS

To carry out this physics program the outgoing $p$ and $\bar{p}$ will be detected in silicon strip detectors (Forward Silicon Trackers, FST). These enable one to move detectors very close ($\approx$ 1 cm) to the circulating beams. The detectors will be in roman pots so that they are in air, and one has accessibility to the detectors which can easily be replaced if necessary. Also there is good screening from electromagnetic pick-up from the beam bunch pulses. The pots are stainless steel vessels which move horizontally in close to the beams when they are stable. There is a 8 mm radius half-cylindrical channel where the beams go. This gives 8 $\sigma$ clearance. If more clearance is needed the pots do not move as far in. There will be a beryllium window (40 mm radius to match the FST) at the front and back of the pots to reduce multiple scattering. We will have three types of detectors: tracking based on silicon strips, and triggering and timing based on scintillator half-discs and fast quartz Cerenkov counters.

### A. Tracking, FST

The detectors must be after dipoles to give acceptance for Feynman $x$, $x_F > 0.90$ over as large a $t, \phi$-coverage as possible, and to measure the $p$ and $\bar{p}$ momenta. At present there is no warm space at CDF (or DØ ) where such detectors could go on the outgoing proton side. CDF already has roman pots on the outgoing antiproton side, and DØ will have them for Run 2A. DØ will also have quadrupole spectrometers on both sides with acceptance for $|t| >\approx 0.6$ GeV$^2$. The spectrometers we propose are superior to the DØ Run 2A set-up in at least four respects:

(1) Acceptance down to $|t| = |t|_{min} \approx 0$ for $0.03 < \xi < 0.10$ on both E and W sides.

(2) A factor $\times$ 20 better spatial resolution (Si strips vs fibers).

(3) Higher and more uniform magnetic fields (dipole vs quadrupole).

(4) Ability to take data at the highest luminosities (using precision timing).

We can make a warm space of $\approx 1.5$ m on the outgoing $p$ side (see below). We already have a lever arm of 2.0 m on the outgoing $\bar{p}$ side, giving $\sigma_{x'} = \sigma_{y'} \approx 3 \times 10^{-6}$ with positioning accuracy of 5 $\mu$m. The Liverpool CDF group have obtained [44] 5 $\mu$m resolution with 32.5 $\mu$m strips in an $r - \phi$ geometry when tilting the detectors by $\approx 6°$. The $p$ and $\bar{p}$ will have traversed 18.8 m of 4.34 Tesla dipoles before entering the detectors.

The acceptance on the $p$ and $\bar{p}$ sides will be very similar but not identical. This is being studied, but we know from Run 1 data (where we had the same situation but with a smaller 2 cm $\times$ 2 cm detector) that the acceptance is 100% at $|t| = |t|_{min}$ for $0.05 < \xi < 0.09$ and out to $|t| \approx 0.6$ GeV$^2$ over most of this range. For $\xi < 0.01$ there is only acceptance for $|t| >\approx 0.6$ GeV$^2$. The lack of acceptance at small $\xi, t$ is good because we are primarily interested in large masses (hence large $\xi$) or large $|t|$, and the trigger rates would be much higher if we accepted small $\xi, t$.

The Liverpool University group are building silicon disc detectors with $r - \phi$ geometry of ideal dimensions for our arms [44]. These are discs (in two 180° half-moons) of outer radius = 40 mm with a circular cut-out for the beams with radius 8 mm. This allows 8 $\sigma$ clearance of the beams in the "fully closed" position, which should be acceptable with the planned improved collimation. If we find that this position gives any background in CDF or DØ or unacceptable rates in our detectors we can always retract. A rad-hard version with 32.5 $\mu$m circular strips and almost-radial ($\approx 5^\circ$ skewed) strips has been built by Micron. We plan to use 4 doublets per arm, with the radial strips oriented to give small angle stereo when combining the two close doublets. This gives a system of $4 \times 8 \times 2048 = 65536$ channels. The chips will be the SVX4 chips as planned for the Run 2B central silicon detectors. We will need 16 128-channel chips per detector, i.e. 576 total so we should make 1024 (5 wafers) to allow for yield. We assume each hybrid handles eight chips so we require 72 hybrids. We need also mini port cards (one per hybrid), junction port cards and cables, DAQ FIB modules and DAQ SRC modules, together with power supplies. The bottom line cost is given in section VIII.

Radiation hardness has been tested on a $p^+n$ 300 $\mu$m detector and doses of $10^{14}$ $p$ cm$^{-2}$ are tolerated. This is at least a factor 10 more than we anticipate in 5 years operation. Signal:noise should be good throughout the lifetime of $>5$ years. Pattern recognition will not be difficult as the track multiplicity in the telescopes will be not much more than 1, and the $\phi$ views will be crossed to give $\phi\pm\approx 5^\circ$ which with $r$ will resolve any ambiguities. The electronics is at the outer periphery where radiation is lower.

There will be a total of four "telescopes" each with two pots: on the inside (S) and outside (N) of the Tevatron and on the outgoing $p$ (E) and $\bar{p}$ (W) sides. Each telescope has precise (we are aiming at 1 $\mu$m positioning reproducibility with 1 $\mu$m position readout) horizontal (N-S) motion to approach the beams. We will also build a 9th roman pot with 4 silicon detectors for use in a test beam.

### *1. Autosurvey*

C. Lindemeyer(PPD), an expert in precision mechanics, has a solution for a mechanical system that will give us 1 $\mu$m positioning accuracy and read-out, *wrt* an external reference. There are also several ways in which the data themselves can be used as a check. The detectors in the front pot will be as widely separated in $z$ as possible ($\approx$ 10 cm) and rigidly mounted together so that they always move as a whole. If we know the position of this front detector unit, fitted tracks in it must point straight back to the detectors in the back pot, as there are no fields in between and the tracks are very stiff ($>$ 900 GeV). So the back detectors' positions can be checked *wrt* the front ones. Elastic scattering events, to be collected continuously at a (prescaled if necessary) rate $\approx$ 10 Hz[1], provide a check on the alignment of the East and West arms, as $t_1$ must equal $t_2$ and $\phi_1 = 180^\circ - \phi_2$. Elastic scattering also provides a check on the relative position of the N and S detector arms. For any selected $t$, the $\phi$-distribution of elastic scattering (which must be flat if acceptance $A = 100\%$) is a direct measure of the acceptance $A(t, \phi, \xi = 0)$. The N and S pots at the "same" location in $z$ will be displaced in $z$ by a pot diameter to reduce any acceptance gap between them. From a mechanical point of view one will be able to overlap them so that some particles can pass through both N and S arms, which checks their relative position at the 1 $\mu$m level. If this cannot be done during standard running, we should be able to do it during occasional short periods at the end of a run.

For low mass exclusive states such as $p\pi^+\pi^-\bar{p}$ we know that $\sum p_x = \sum p_y = \sum p_z = 0$ and $\sum E = \sqrt{s}$. The first two constraints are especially powerful: plots of these quantities (the sum is over the $p, \bar{p}$ and the central charged particles) will show a narrow peak centered at 0 if the alignment is perfect. There will be events outside this peak due to cases where one or more particles have not been detected. The $p_T$ resolution of the pot tracks is $\approx$ 20 MeV, and the resolution from the central trackers on $\sum p_T$ of a 2- or 4-particle state in the low mass ($\approx$ 1 GeV) region is similar.

For higher masses ($\xi \geq 0.03$) the $|t| = |t|_{min} = |2[m_p^2 - (E_{in}E_{out} - p_{in}p_{out})]|$ point falls inside the trackers. Plotting the data as a function of $t$ for fixed $\xi$ one can check the position of this sharp $|t|_{min}$ edge. Another way of seeing the same thing is to plot the $x, y$ distribution of hits in a detector for fixed $\xi$, and observe the position of the point of maximum density. When the alignment is satisfactory a plot of the $\phi$ distribution for fixed $\xi$ and fixed $t$ gives $A(\phi)$.

---

[1]Note that, recognizing these events to be elastic at a Level 2 trigger, only the VFTD detectors need be read out, so the events are *very* small.

## B. Timing and Trigger counters

Each of the four arms will have a thin (2.5 mm) scintillation counter, $S_i$, read out via a twisted strip light guide to a PMT at the front, and a 5-element quartz Cerenkov hodoscope, $Q_{ij}$, at the back. The "arm trigger" is the coincidence $S_i Q_i$ where $Q_i$ is the OR of the 5 counters $j$.

### *1. Scintillation Counters*

At the front of each arm will be a thin (2.5 mm) plastic scintillator of identical size and shape as the silicon half-discs. Around the outer edge will be attached seven 18 mm wide isochronous twisted strips, which will be brought together in a 18 mm $\times$ 17.5 mm block mounted on the photocathode of a Hamamatsu R5800U PMT. With 20% light collection efficiency and a 20% Quantum Efficiency we expect $\approx 200$ photoelectrons per $p$ or $\bar{p}$. The multiple scattering of a 900 GeV proton in this counter is $\sigma \approx 3.5$ $\mu$rad which becomes non-negligible (so we may try thinner counters).

### *2. Fast Timing Cerenkovs (FTC)*

With multiple interactions in a bunch crossing a background can come from two single diffractive collisions, one producing the $p$ and the other the $\bar{p}$. One way of reducing this is to require longitudinal momentum balance $\sum p_z = 0$. However this "pile-up" can be further reduced by a factor $\approx 25$ by the quartz Cerenkov counters FTC (Fast Timing Cerenkovs) which have excellent timing resolution. Quartz is radiation hard and has good transmission in the UV. One can achieve $\delta t = 30$ ps timing resolution on the $p$ and $\bar{p}$, much better than the ($\approx 1$ ns) spread between random concidences. The sum of the $p$ and $\bar{p}$ times referred with respect to the interaction time as measured by the central TOF barrel is a constant for genuine coincidences. Their difference $\Delta t$ is a measure of $z_{\circ}$ of the interaction at the level of 1 cm (for $\delta t = 30$ ps). CDF has a Time-of-Flight barrel of 216 counters in $|\eta| < 0.75$ with resolution $\approx 100$ ps per particle, or $< 50$ ps on a $b$-jet. One can do a global timing fit between the $p$, the $\bar{p}$ and the central particles (if these are in the TOF barrel). We need four of these Fast Timing Cerenkov FTC detectors, one for each arm (plus a 5th for the test beam pot). Each one consists of five polished quartz bars 18 mm $\times$ 18 mm $\times$ 30 mm or 40 mm (horizontal). There are no light guides; the quartz bars are directly glued to the PMT window. The particles traverse 18 mm and the number of photoelectrons is expected to be approximately

$$N_{pe} \approx \frac{1}{2} \times 90 \times L(cm)sin^2(\theta) \approx 40$$

where the factor $\frac{1}{2}$ is put in because we ignore light emitted away from the phototube, and the refractive index of quartz is $n = 1.458$ so the Cerenkov angle is $\theta = 46.7^\circ$. Light does not emerge through the front or back surfaces because it is totally internally reflected; reflectors are put on the top and bottom surfaces, and the face opposite the PMT may be specially treated (made absorbing or reflecting depending on what gives the best time resolution with enough light). There are 5 blocks in an FTC detector and they overlap in $y$ by 2.5 mm, with a displacement in $z$. A few percent of the particles will be measured in two independent blocks, which gives a monitor of the time resolution (as well as a factor $\sqrt{2}$ better time resolution for those tracks!). We will be able to apply off-line time-slewing corrections if the time measurement is correlated with the pulse-height. As we know precisely the track position in $(x,y)$ we can also apply a correction for that. Our present choice of photomultiplier is a Hamamatsu R5900U which has a square photocathode of 18 mm $\times$ 18 mm and is less than 30 mm deep (without the socket). We will use the same PMT for the scintillation counters.

### C. Modifications to the Tevatron

At present there is no warm space for the detectors on the outgoing $p$ side, and some modifications will have to be made to the Tevatron to generate such a space. Fortunately the Q1 quadrupole at B-11 is no longer being used and it can be removed[2], releasing 1.850 m of space. The dipoles B11-2,3 and 4 will be moved into this space, making free space between the 3rd and 4th dipoles (B11-4 and B11-5) in B11, which is where we want to put the roman pots. One scenario which gains more space (desirable since about half of the 1.85 m is used up by the ends of the bypass, two vacuum valves, flanges and bellows) is to replace both the Q1 spool and the R spool with a standard Tevatron H spool and a Collins quad adapter. In this case we would open up 280 cm, providing more than 180 cm for detectors. Other dipoles have to move to balance this change. A nice solution proposed by P. Bagley is to shift six dipoles in B16 and B17 by half as much in the opposite direction, away from B0. This is made possible by replacing the D-spool at B18 with a short B-spool. One can also remove a 40 cm spacer in B17-5. A consequence is that the section of Tevatron from B11-5 to B16 spool BQ9 must be moved to the radial inside by 3.6 cm = 1.4". From a visual inspection we [9] have ascertained that these modifications can be done without any difficult problems. Some pipe extensions will have to be made but there is essentially no cable work other than remounting a section of cable tray. The main issue seems to be labor. We believe that this can be done in a two month shutdown. New cryogenic bypasses are needed for B11 and B16. This modification reduces the circumference of the Tevatron by about 6.2 mm and so reduces the radius by about 1.0 mm. At present the radius of the Tevatron is about 6.3 mm too large (compared with design and with the best match to the Main Injector). So this is a small move in the right direction. This does not change the working point of the machine and it is not expected to add significant time to the recommissioning after the shut down.

[2]Both Q1 have already been removed at DØ .

In this scheme (unlike in some alternatives we have considered) there is no displacement of CDF or the straight section containing CDF, with its low-$\beta$ quadrupoles and electrostatic separators. Also the A-side (outgoing $\bar{p}$), which contains the existing roman pots, is left untouched. These are major advantages; it is expected that repositioning CDF (by 45 mm in the simplest scheme) would be expensive and take a shutdown considerably longer than 2 months. On the A-side we simply replace the existing roman pots with the new ones.

### D. Running Conditions and Triggers

We shall not require any special running conditions; we use the normal low-$\beta$ tune. (Of course we will need a little beam time to do commissioning tests.)

We will record the information in these detectors for every CDF event. Then one can look for forward protons in any physics process under study. However we will need forward triggers, requiring a forward track in both arms. The arm trigger is based on a coincidence between the scintillator and the Cerenkov counter. This will be at Level 1, and the rate at L = $10^{31}$ cm$^{-2}$ s$^{-1}$ is expected to be $\approx$ 5 KHz. At Level 2 a trigger processor will calculate tracks and find the $MM$, and recognize elastic events for special treatment (writing only the pot information, when the central detector appears to be empty).

We want the maximum integrated luminosity and we aim to be able to take good data with the maximum luminosity the Tevatron can deliver. We discuss triggers more specifically after presenting the physics program.

## III. EXCLUSIVE HIGGS BOSON PRODUCTION

If the Higgs is produced with a large enough cross section in the exclusive [3] reaction $p\bar{p} \to pH\bar{p}$ it will give rise to a peak at $M_H$ in the missing mass spectrum. High resolution makes a Higgs search feasible over the full mass range 110-180 GeV (we now know from LEP that $M_H > 113.5$ GeV) at the Tevatron with 15 fb$^{-1}$ as hoped for in Run 2. Up to about 130-140 GeV the $b\bar{b}$ and $\tau^+\tau^-$ modes can be used, above 135 GeV the $WW^*$ mode takes over and above 160 GeV the $WW$ mode dominates. For the $\tau$-pairs and using only the leptonic decay modes of the $W$-pair the signal is extremely clean because, unlike generic lepton pair production, there are no hadrons at the primary vertex. Thus a 160 GeV Higgs can appear as a final state with $p_3 + l_1^+ + l_2^- + \not{E}_T + p_4$ with no other particles on the $l_1 l_2$ vertex. Such events should be easily recognizable even with many interactions in a bunch crossing, using knowledge of $z_\circ$ from the precision timing. At the high end of the mass range the mass resolution becomes better than the width of the Higgs, which could then be measured. The ratio of events in the channels $b\bar{b}, \tau^+\tau^-$ and $W^+W^-$ can demonstrate the coupling of the Higgs to mass. Production and decay angular distributions can demonstrate that it is a scalar. The visibility of this signal depends on the exclusive cross section. Some theoretical calculations are very encouraging, while others claim that the cross section should be too low for the Tevatron, but perhaps not too low for the LHC. There are differences of more than a factor 100 in the predictions. The relevant diagrams are shown in Fig 1. The main problem in calculating these is the soft non-perturbative nature of the second gluon (or more gluons), for which we do not have a well accepted theory, and the difficulty in estimating the probability that no other particles will be emitted. If $H$ is not seen in this proposed experiment that will rule out some models. If it *is* seen, that might not only be a discovery but it will provide the best way at hadron colliders of measuring $M_H$ with an uncertainty of $\approx$ 250 MeV *per event*.

[3] Much of this section is based on reference [10].

FIGURES

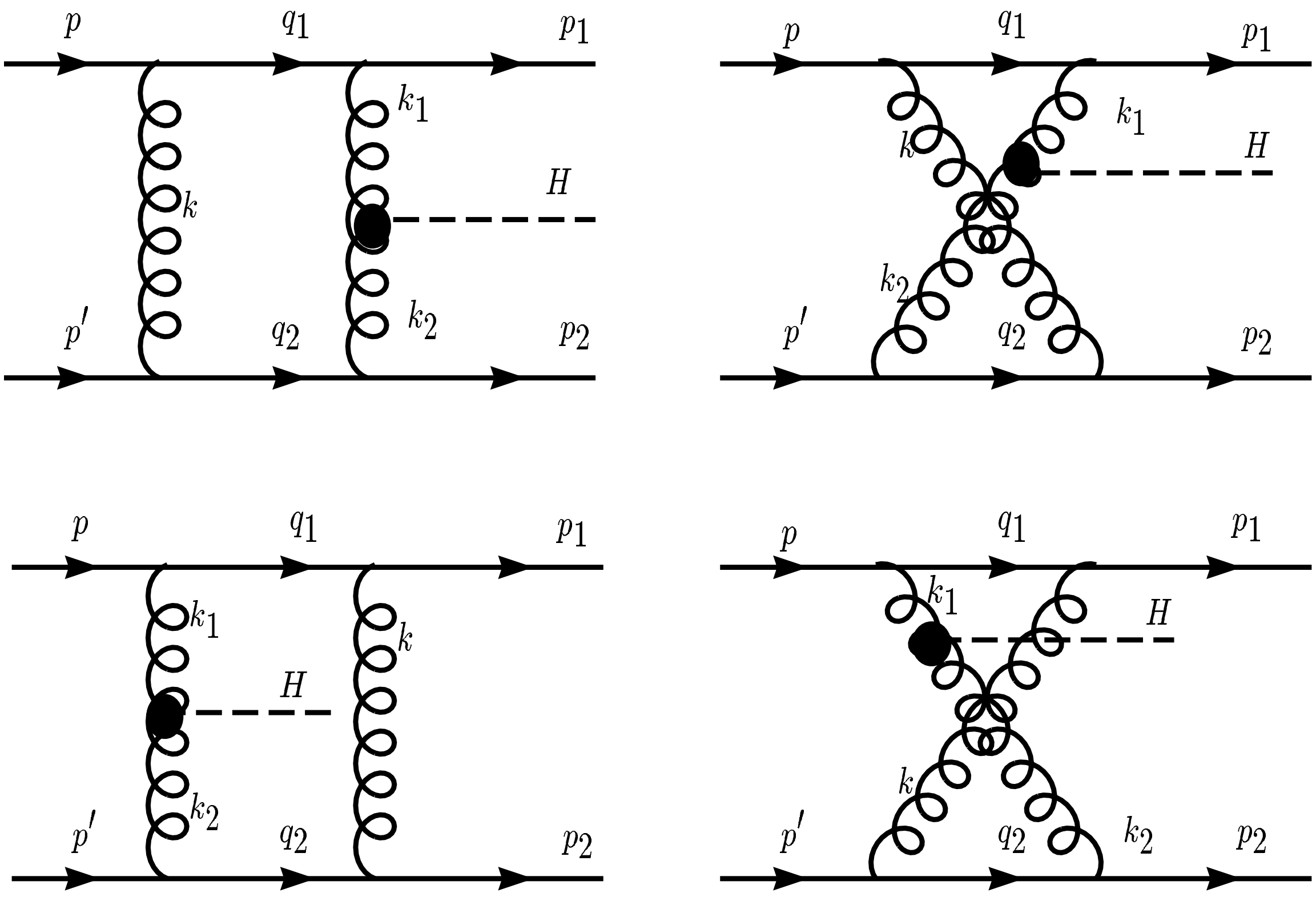

FIG. 1. Exclusive Higgs production diagrams ($CH$).

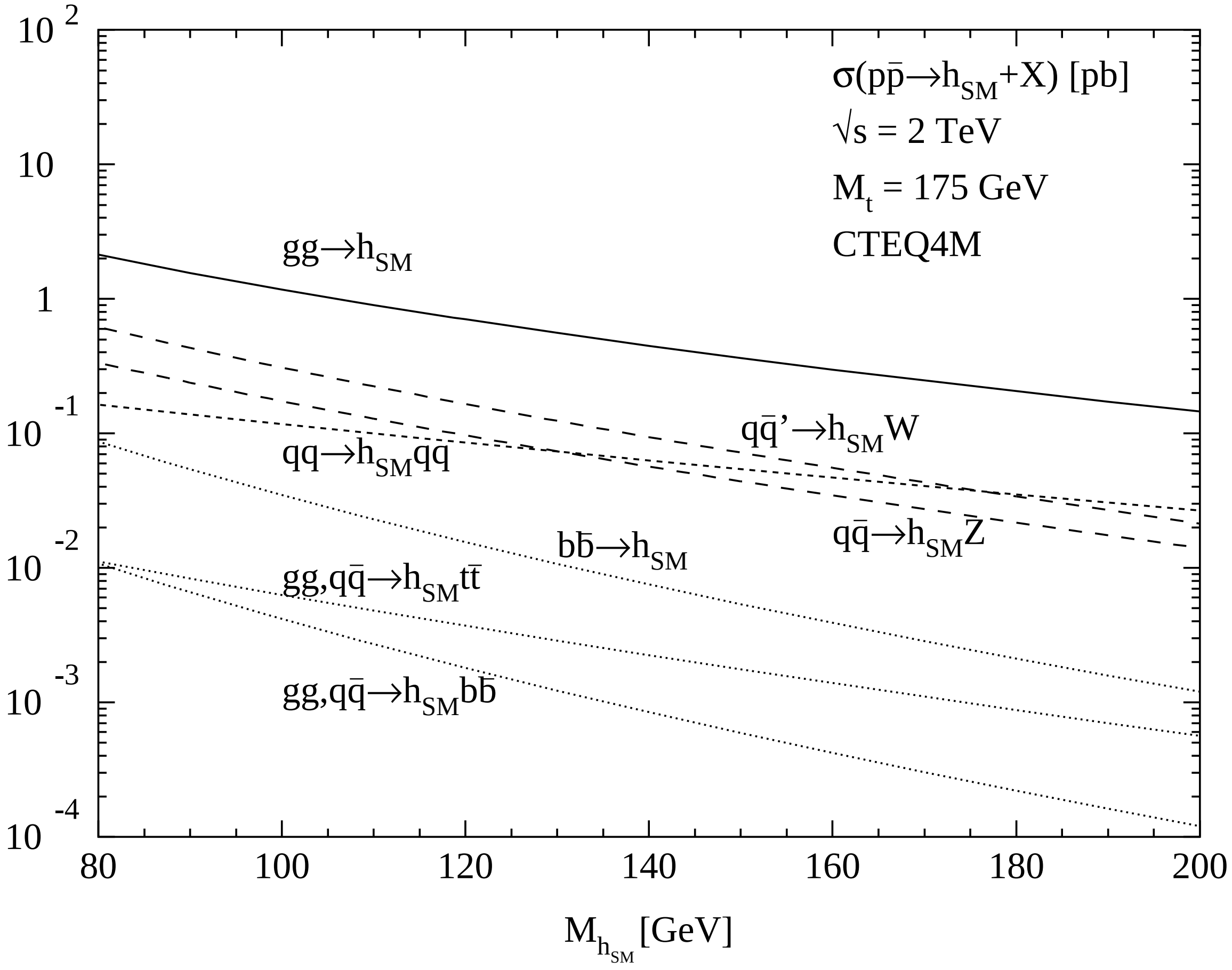

FIG. 2. Inclusive Higgs production cross sections vs $M_H$.

The predominant mode for Higgs production at hadron colliders (see Fig.2) is $gg$-fusion [11,12] through a virtual top quark loop (in the 3-generation SM). The dominant decay mode up to 135 GeV is to $b\bar{b}$ (Fig.3), above which the $WW^*$ mode becomes increasingly important until $M_H > 2M_W$ (160 GeV) when both $W$ are real. By 200 GeV the $ZZ$ mode has grown to 26%. The $\tau^+\tau^-$ mode decreases from 7.3% at 115 GeV to about 2% at 150 GeV. The intrinsic width of a Higgs over this mass region rises, from only 3 MeV at $M_H$ = 120 GeV, to 16 MeV at $M_H$ = 150 GeV, to 650 MeV at $M_H$ = 180 GeV (Fig.4) [11]. Mass resolution is therefore crucial in increasing the signal:background $S : B$ ratio.

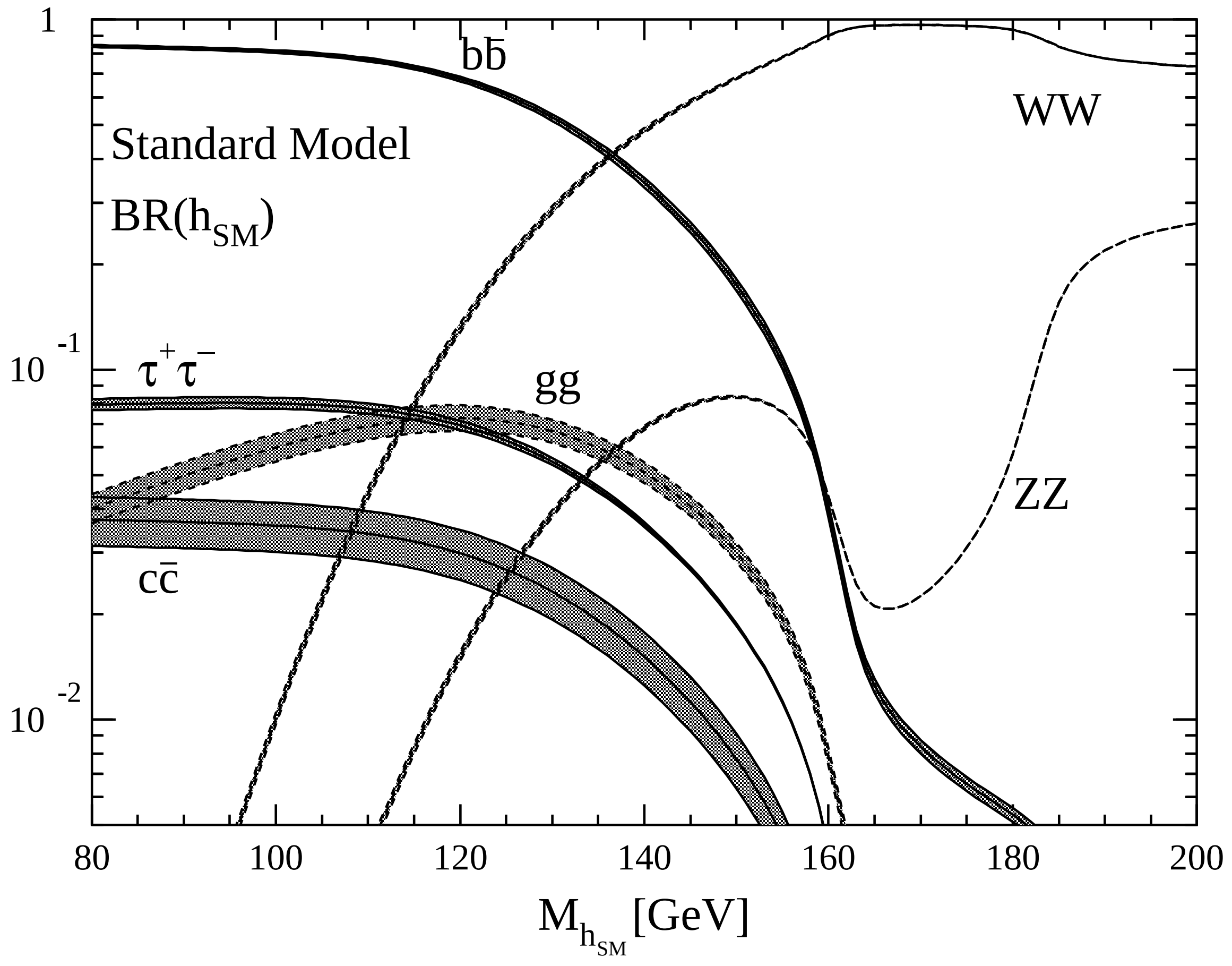

FIG. 3. SM Branching fractions as a function of $M_H$.

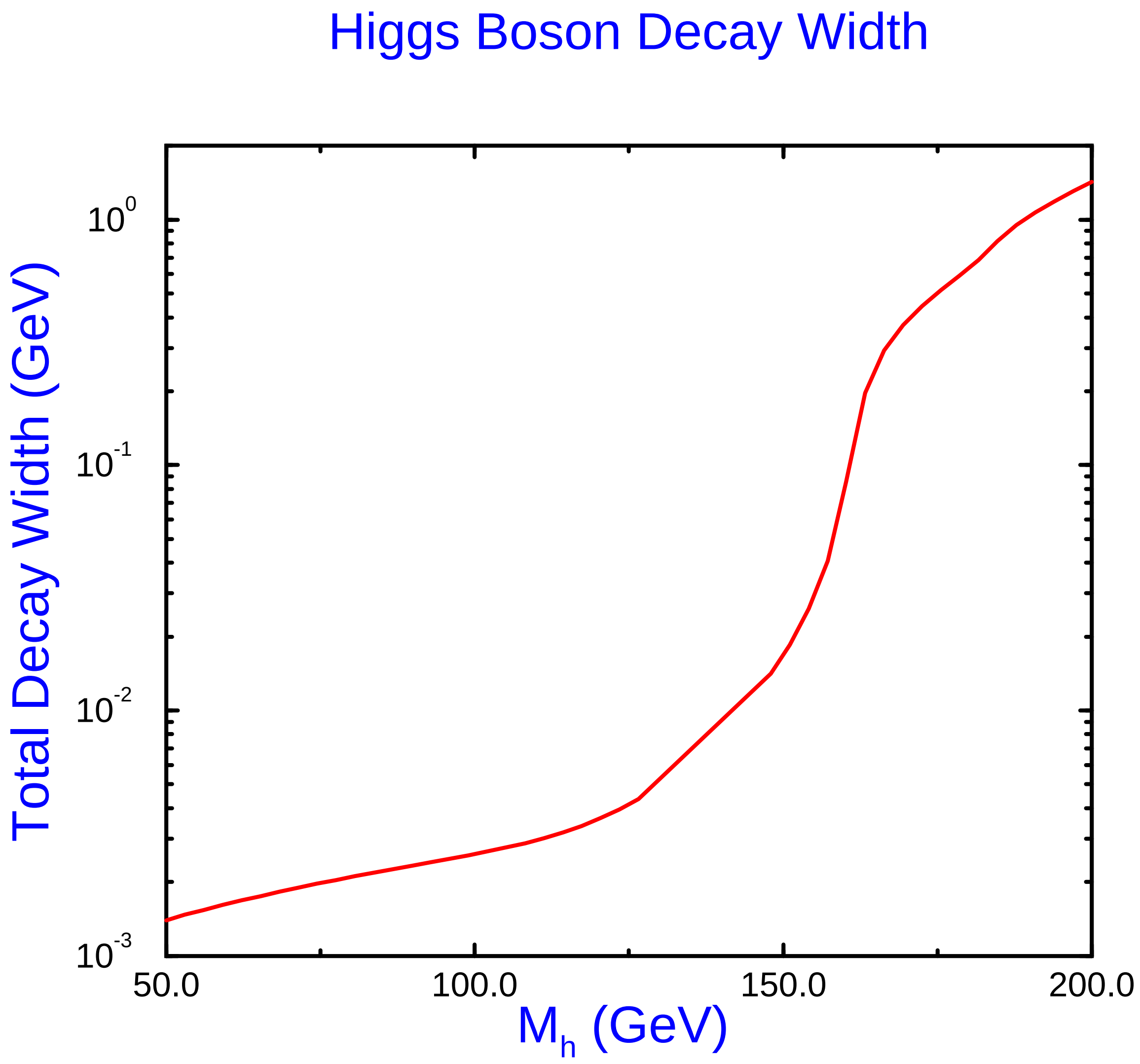

FIG. 4. SM Higgs width as a function of its mass.

One has generally supposed that the observation of the Higgs in the intermediate mass region 110 GeV to 130 GeV in hadron collisions is impossible because of the small $S : B$, *unless* one selects the relatively rare cases where it is produced in association with a massive particle $(W, Z, t)$ or where it decays to $\gamma\gamma$ (branching fraction $\approx 2 \times 10^{-3}$), where much better mass resolution can be obtained than for any other final state. A high price has to be paid for these requirements. In 15 fb$^{-1}$ we expect more than 10,000 120 GeV $H$ to be produced and 70% of them decay to $b\bar{b}$. However the mass resolution in reconstructing a $b\bar{b}$ di-jet is about 10 GeV - 15 GeV, and the QCD background is indeed overwhelming when the signal is so spread out. Using the missing mass method that we propose, the resolution is improved to 250 MeV, increasing the $S:B$ by a factor $\approx$ 40 - 60. The method works not only for $b\bar{b}$ Higgs decays but also for $\tau^+\tau^-$, $W^+W^-$ and $ZZ$ decays, and the number of neutrinos in the final state is irrelevant for the mass resolution. In fact neutrinos are turned into an asset as they give missing $E_T$ ($\not{E}_T$) which is a positive signature and can be used in a trigger.

The visibility of a signal will depend on the spread in the beam momenta $\frac{\delta p}{p}$ and the measurement error on $p_3$ and $p_4$. Any overall scale factor such as would come e.g. from uncertainty in the magnetic fields in the Tevatron only affects the central value, i.e. $M_H$ if a signal is seen. The momentum spread of the incoming beams [13] is 1.0 $\times 10^{-4}$ at the beginning of a store and rises to about 1.6 $\times 10^{-4}$ after 20 hours of collisions. Their divergence is $\approx$ 100 $\mu$rad. These two effects contribute about equally to $\sigma_{MM}$. The position of the interaction point $x_\circ, y_\circ, z_\circ$ will be reconstructed in the SVX with $\sigma \approx 4$ $\mu$m, 4 $\mu$m and 10 $\mu$m respectively for central $b\bar{b}$ jets, and about a factor two worse [4] for $l^+l^-$ final states. The outgoing $p$ and $\bar{p}$ tracks will be measured using eight layers of silicon detectors ($R, \phi$',$\phi$") giving $\sigma_x = \sigma_y \approx 5$ $\mu$m over $\approx$ 1.0 (2.0) m, thus $\sigma_{x'} = \sigma_{y'} \approx 3(1.5) \times 10^{-6}$. If $\sqrt{s}$ is the center of mass energy (1.96 TeV) and the outgoing scattered beam particles have lost fractions $\xi_1, \xi_2$ of their incident momenta, we have approximately $MM^2 = \xi_1 \xi_2 s$. The spread in the reconstructed missing mass, $\delta_{MM}$ is a combination of the relative spread $\frac{\delta p_b}{p_b}$ in the beam particles' momenta $p_b$ and their divergence, and the resolution of the "dipole spectrometers" which use the primary interaction point and the outgoing tracks. With the above parameters this is $\approx$ 250 MeV, independent of $MM$.

We note that this method is not limited to Higgs searches but would be sensitive to any relatively narrow massive objects with vacuum quantum numbers.

The visibility of the Higgs by this technique clearly depends on the size of the exclusive cross section. The mechanism $gg \to H$ normally leaves the $p$ and $\bar{p}$ in color-octet states and color strings fill rapidity with hadrons. However some fraction of the time one or more additional gluons can be exchanged which neutralize (in a color sense) the $p$ and $\bar{p}$ and can even leave them in their ground state (see Fig.1). In

---

[4] We assume both leptons are tracked in the silicon vertex detectors.

Regge theory this is the double pomeron exchange ($DIPE$) process Several attempts have been made to calculate this cross section.

In 1990 Schäfer, Nachtmann and Schöpf [14] considered diffractive Higgs production at the LHC and SSC, concluding that the cross sections for the exclusive process could not be reliably predicted.

Müller and Schramm [15] made a calculation, also for nucleus-nucleus collisions, and concluded that the exclusive process is immeasurably small. Basically this is because they take the pomeron to be an extended object and it is very difficult to "localize" pomerons to order $M_H^{-1}$. We have since learnt that this is not a valid picture for hard interactions. If it were true we would never find large rapidity gaps between balancing high $E_T$ jets, which actually occur [16] at the level of 1%. A more realistic picture is that a hard $gg$-interaction occurs, and the color removed from the $p$ and $\bar{p}$ is neutralized on a much longer time scale by one or more additional soft gluon exchanges. The probability of this neutralization happening, with color octet gluons, is $\approx \frac{1}{64}$ with an additional factor called the *rapidity gap survival probability.*

In 1991 Bialas and Landshoff [17] calculated from Regge theory that about 1% of all Higgs events may have the $p$ and $\bar{p}$ in the $DIPE$ region of $x_F \approx 0.95$.

In 1994 Lu and Milana [18] obtained an estimate "well below what is likely to be experimentally feasible".

In 1995 Cudell and Hernandez [19] made a lowest order QCD calculation with the non-perturbative form factors of the proton tuned to reproduce elastic and soft diffractive cross section measurements. They presented the exclusive production cross section as a function of $M_H$ up to 150 GeV at $\sqrt{s} = 1.8$ TeV (see Fig.5).

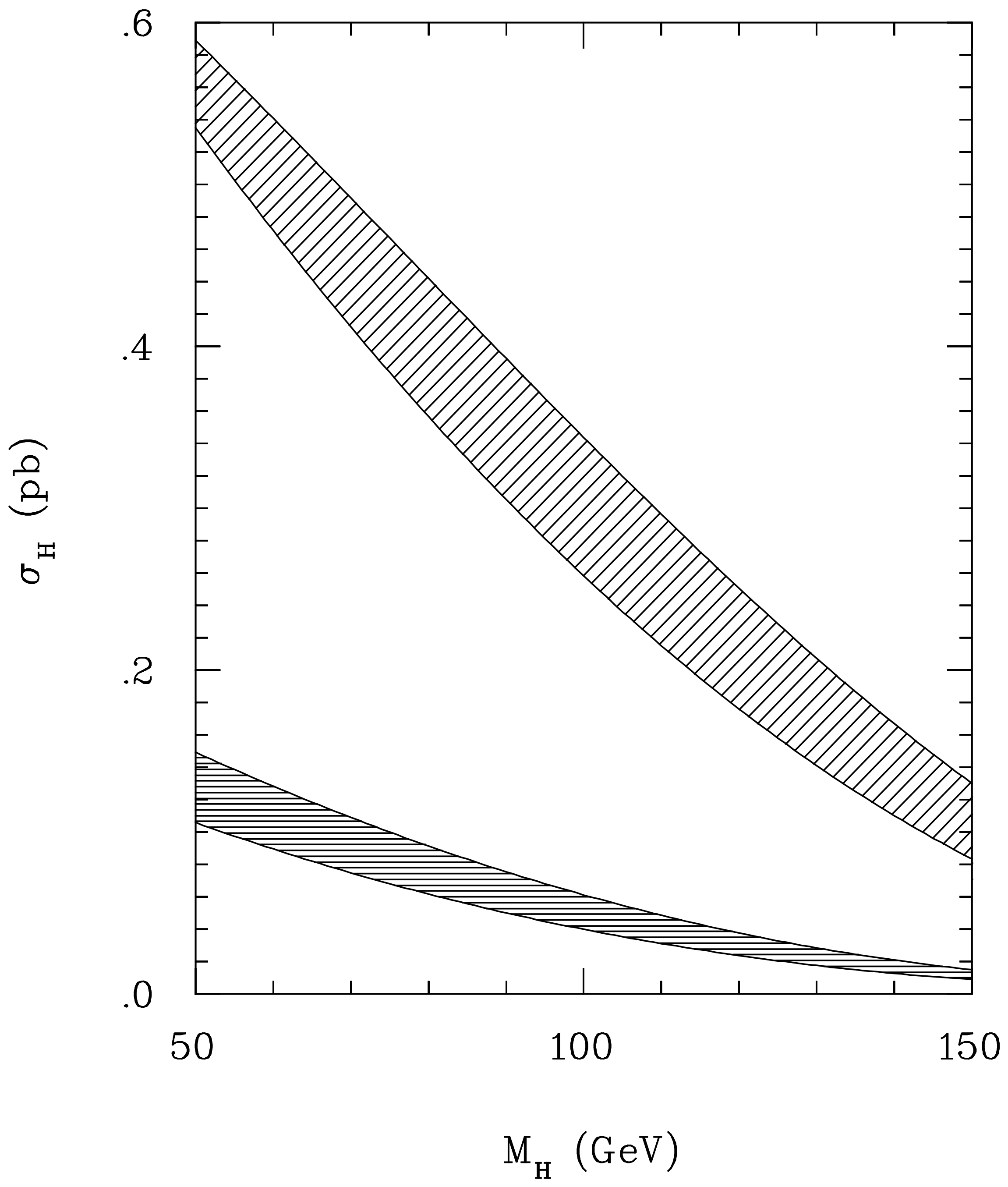

FIG. 5. $CH$ exclusive cross-section $p\bar{p} \to pH\bar{p}$ (lower band). The upper band is when the $p$ and/or $\bar{p}$ are allowed to diffractively dissociate.

They found a cross section decreasing slowly with $M_H$ from 38 fb at 115 GeV, 13.5 fb at 150 GeV and, by extrapolation, 6.0(1.5) fb at 170(200) GeV (all within a factor two). The total Higgs production cross section by the dominant $gg$-fusion mechanism is [12] 800 fb, 364 fb, 247 (145) fb respectively so the exclusive fraction decreases from 5% to about 1% over this mass range, even higher than the Bialas and Landshoff estimate. However there are issues of "rapidity gap survival probability", "pomeron flux renormalization" [20], shadowing effects, initial and final state interactions etc. These effects are (not necessarily different) ways of explaining why diffractive cross sections in hadron-hadron (but not $ep$) collisions are about an order of magnitude lower at high $\sqrt{s}$ than naïve Regge expectations. There are two recent calculations. Khoze, Martin and Ryskin [21] find $\sigma(pp \to pHp) = 0.06$ fb for $M_H = 120$ GeV at $\sqrt{s} = 2$ TeV if the probability $S^2_{spect}$ not to have extra rescattering in the interaction is $S^2_{spect} = 0.05$. This is too low for the Tevatron.

Kharzeev and Levin [22] find much higher values of 19 - 140 fb for $M_H = 100$ GeV at the Tevatron, but they do not present the $M_H$-dependence. Differences from Ref [21] are associated with the treatment of the exchanged gluons and final state bremsstrahlung.

For the channels $H \to \tau^+\tau^-$ and $H \to WW^{(*)}$ with both $W$ decaying leptonically, we can allow additional hadrons with 4-momenta $p_i$ as long as they are measured in the central detectors. Then:

$$MM^2 = (p_1 + p_2 - p_3 - p_4 - \Sigma p_i)^2$$

This will increase the cross section significantly. In a recent review [23] Landshoff reiterates his view that the exclusive production cross section should be large. Although there are large differences in the theoretical predictions, we shall show that the higher predictions allow a Higgs discovery at the Tevatron over the full mass range from 110 GeV to 180 GeV. The 2-gap survival probability in the central Higgs case is not necessarily the square of the 1-gap survival probability, because the Higgs is colorless and its decay products (if it is light, with $\Gamma_H < 5$ MeV say) emerge much later than the formation time of all the other hadrons in the event. If it is heavy ($M_H > \approx 150$ GeV) it decays faster, on hadronization time scales, but we look at dilepton final states which do not couple to gluons. The $t\bar{t}$-loop is too small to interact with soft gluons. One should perhaps rather think of the "non-interacting" Higgs as being produced in the middle of one long (15 units) rapidity gap. The situation is reminiscent of rapidity gap survival in $ep$ collisions, where the electron and the virtual photon do not interact with soft gluons, and in this case the gap probability ($\approx$ 10 %) is a factor $\approx 10$ higher than in hard $p\bar{p}$ collisions. In our case we may also have a 2-gap probability $\approx (10\ \%)^2 = 1$ %, rather than $(1\ \%)^2 = 10^{-4}$ which is the approximate level of $DI\!PE$ in the $gg \to b\bar{b}$ background.

We take the Cudell and Hernandez ($CH$) prediction as our benchmark, ignoring any gain from the $\sqrt{s}$ increase from 1.8 TeV to 1.96 TeV and noting that the $CH$ estimate has a factor $\approx 2$ uncertainty. The $CH$ predictions for the Standard Model Higgs are neither the most "optimistic" nor the most "pessimistic" and we take them as an example[5]. We consider signals and backgrounds, first for $b\bar{b}$, then for $\tau^+\tau^-$ and then for $WW^{(*)}$ using only the leptonic decays $l_1^+ l_2^- \nu\bar{\nu}$. We also consider $W^+W^-$ decaying to $l^\pm \nu jj$. Table 1 shows a compilation of results.

[5] $CH$ do not include rescattering suppression or Sudakov effects which can reduce the exclusive cross section.

## TABLES

| $M_H$ (GeV) | $\sigma(CH)$ (fb) | Mode | BR | $\sigma$.BR.BR (fb) | Events $15\text{fb}^{-1}$ | Background /250 MeV |
|---|---|---|---|---|---|---|
| 115 | 38 | $b\bar{b}$ | 0.730 | 27.7 | 208 | $< 23.4$ |
| | | $\tau^+\tau^-$ | 0.073 | 2.8 | 21 | $< 0.1$ |
| 130 | 25 | $b\bar{b}$ | 0.525 | 13.1 | 96 | $< 7.5$ |
| | | $\tau^+\tau^-$ | 0.054 | 1.35 | 10.0 | $< 0.1$ |
| | | $WW^*$ | 0.289 | 0.72 | 5.4 | $\ll 1$ |
| 150 | 13.5 | $WW^*$ | 0.685 | 0.93 | 7.0 | $\ll 1$ |
| 170 | 6.0 | $W^+W^-$ | 0.996 | 0.58 | 4.3 | $\ll 1$ |
| 180 | 3.5 | $W^+W^-$ | 0.935 | 0.34 | 2.5 | $\ll 1$ |
| 170 | 6.0 | $W(l\nu)W(jj)$ | 0.996 | 2.49 | 18.5 | $\ll 1$ |
| 180 | 3.5 | $W(l\nu)W(jj)$ | 0.935 | 1.47 | 11.1 | $\ll 1$ |

TABLE I. For various Higgs masses, the exclusive production cross section according to Cudell and Hernandez at 1.8 TeV. Column 5 shows the cross section $\times$ branching fractions either to two b-jets or to two charged leptons, or, for the last two rows, one $W$ decaying leptonically and one hadronically. A factor 0.5 has been applied to events and background for acceptance/efficiency.

### A. $H \to b\bar{b}$

For the $b\bar{b}$ dijet background we take CDF's published cross section [24] $\frac{d\sigma}{dM_{JJ}}$ for two b-tagged jets, which starts at 150 GeV (see Fig.6), and extrapolate the fit to the data (which is a factor 2-3 higher than the PYTHIA prediction) down to 115(130) GeV finding 125(40) pb/GeV (in $|\eta| < 2.0, |cos(\theta^*)| < 2/3$).

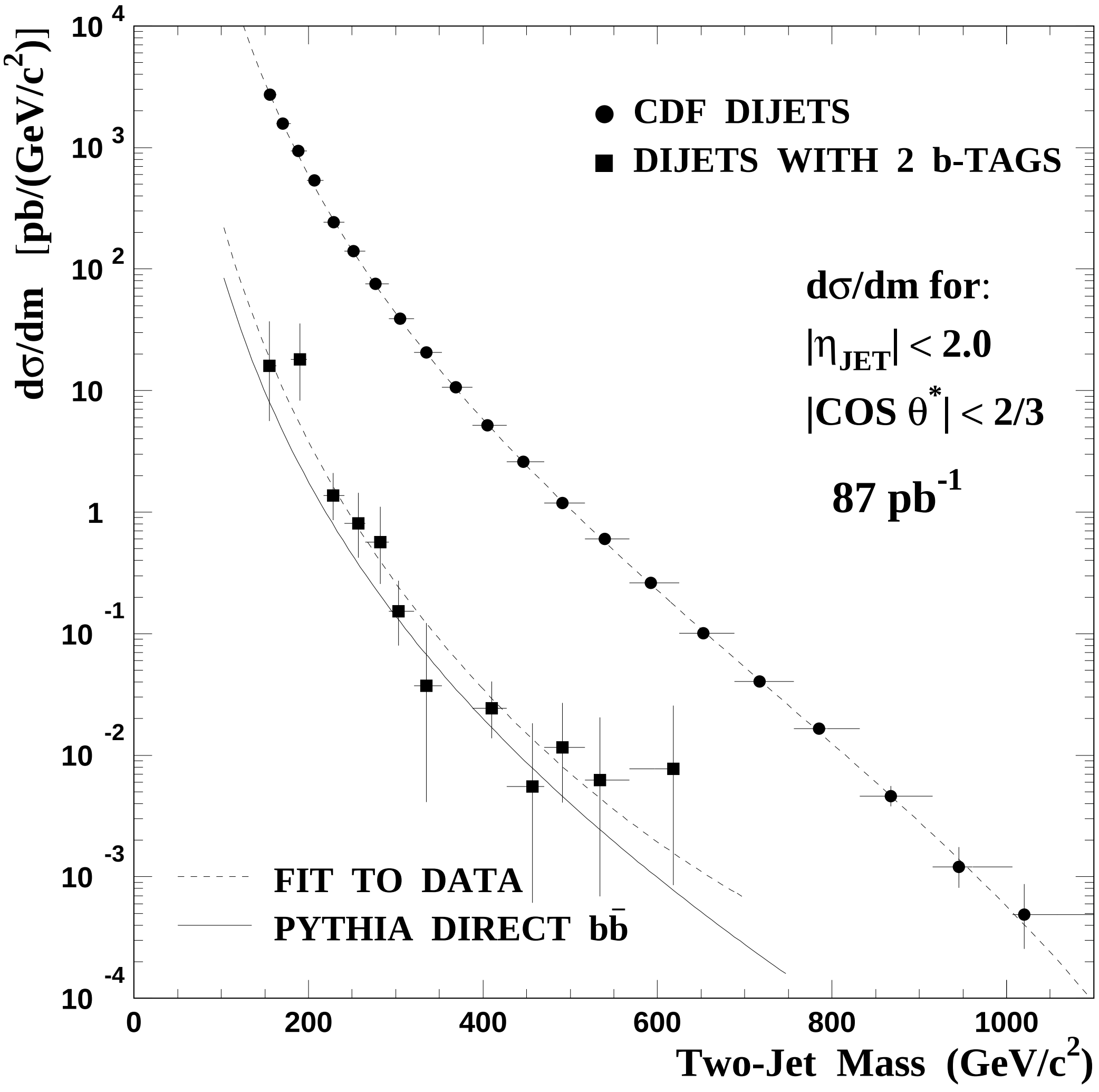

FIG. 6. Dijet and $b\bar{b}$ dijet cross section vs $M_{JJ}$ (CDF).

From our other $DI\!PE$ studies, of lower mass dijets [6], we expect that less than $10^{-4}$ of these are $DI\!PE$ ($p\bar{p} \to p+b\bar{b}+\bar{p}$), where $+$ represents a rapidity gap exceeding about 3 units, assuming this fraction is not $E_T$-dependent. If the fraction is smaller, so much the better. That gives 5(1) fb per 250 MeV bin, to be compared with a signal of around 45(25) fb [19]. With 15 fb$^{-1}$ and assuming 50% acceptance for both signal and background we have 260(96) events (see Table 1) on a background of 37.5(7.5). Even if the $CH$ predictions are too high by an order of magnitude these signals are 4.2(3.5)$\sigma$. We have not put in a factor for b-tagging efficiency (which affects the signal and the background the same way apart from differences in the angular distributions); it was about 35% per jet in Run 1 at $M_{JJ} = 200$ GeV. It will be higher in Run 2 with more silicon coverage and at smaller masses. For $b\bar{b}$ dijet identification we will cut on a combination such as the product $B_1 B_2$ where $B_i$ is the probability of jet $i$ being a $b$-jet. We have put in an acceptance of 50% for the signal and background, assuming the $|t|$-distribution is as expected for high mass $DI\!PE$, $\approx e^{b(t_1+t_2)}$ with $b \approx 4$. The $H$ has small $p_T$ ($<\approx$ 2 GeV) and being heavy is mostly produced with small rapidity. When the difference in $p_z$ of the forward $p$ and $\bar{p}$ is $< 50$ GeV, as it must be for our acceptance, $y_H < 0.41$ (0.28) for $M_H = 120$ (180) GeV. It decays isotropically. Fig.7 shows the signal for $H(130)$ under the above assumptions.

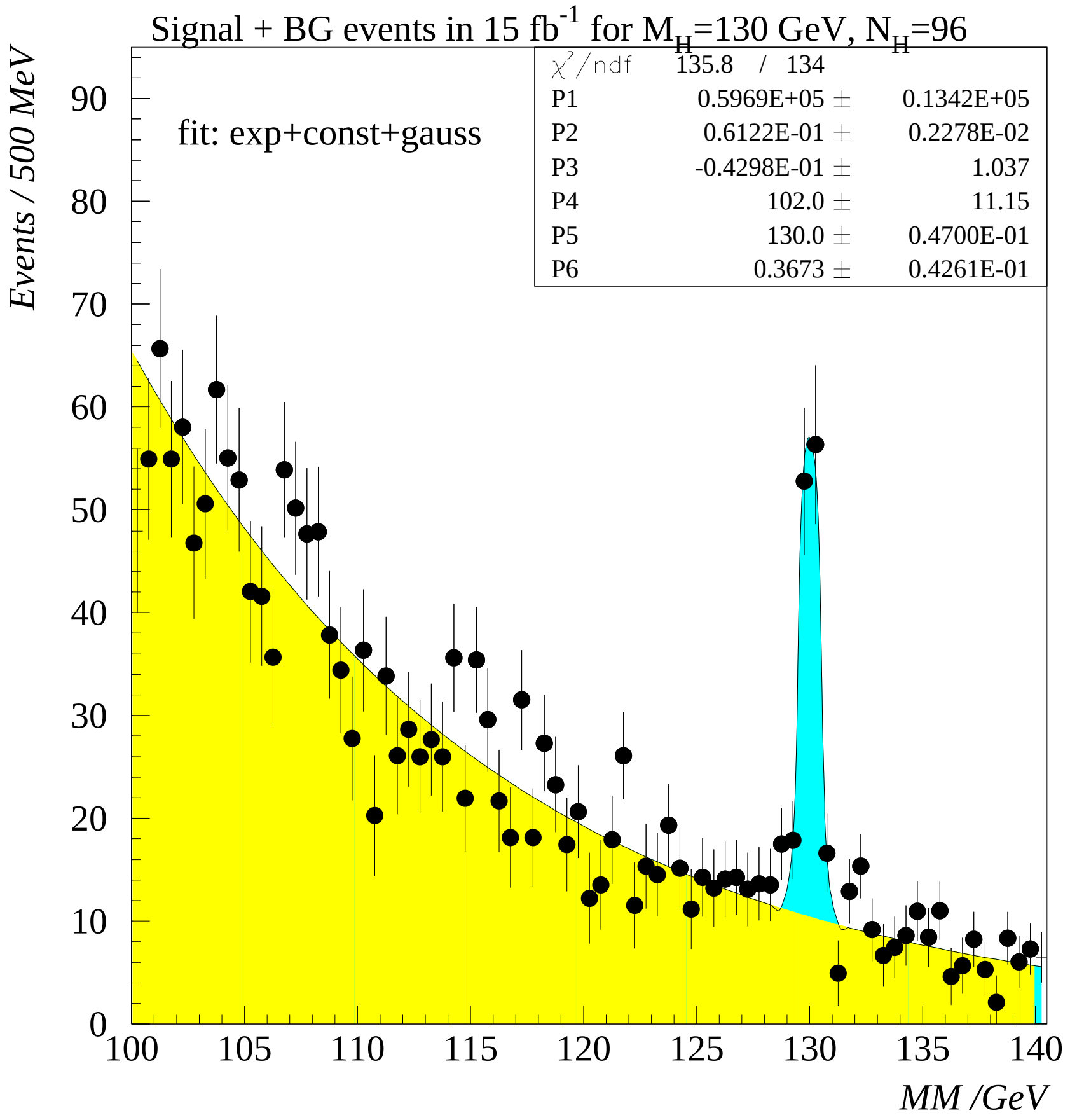

FIG. 7. Simulation of Higgs signal and background in $b\bar{b}$ channel in 15 fb$^{-1}$ according to $CH$, with 50% acceptance assumed.

The $S:B$ ratio rises with $M_H$ in this mass region 115-130 GeV, because the Higgs production cross section falls less steeply than the QCD backgrounds as the top loop becomes more real and the Higgs couples to mass. As $M_H$ increases beyond 130 GeV the branching fraction for $H \to b\bar{b}$ drops rapidly.

### B. $H \to \tau^+\tau^-$

The Higgs branching fraction to $\tau^+\tau^-$ decreases from 7.3% at 115 GeV to 5.4% at 130 GeV, as the $WW^*$ mode grows in competition. Backgrounds to the proposed search could come from normal Drell-Yan $(DY)/Z$ production together with 0,1, or 2 associated high-$x_F$ tracks; in the first two cases leading (anti-)protons come from different events (pile-up); we discussed ways of minimizing this in section IIB. In the third case the events look like continuum $DI\!PE$ production of $DY$ pairs, together with associated particles. Recent CDF studies [5,6] of diffractive jet production at low $E_T$ have found a breakdown of factorization for jet production in the sense that $\frac{\sigma_{DI\!PE}}{\sigma_{SD}} \approx 5 \times \frac{\sigma_{SD}}{\sigma_{ND}}$. Let us assume this fraction is the same for high-mass $DY$, and then assume factorization break-down by the same factor 5 for high mass $DY$. Then $DI\!PE$ production of high mass $DY$ is at the relative level of $5.10^{-4}$. From a CDF study [25] of high mass $e^+e^-$ and $\mu^+\mu^-$ we infer that $\frac{d\sigma}{dM}$ for the region 115-130 GeV is $100 \pm 40$ fb GeV$^{-1}$. Therefore the cross section for $p\bar{p} \rightarrow p + \mu^+\mu^- X + \bar{p}$, where $X$ represents additional associated hadrons, $n_{ass}$ of which are charged tracks, is expected to be about 100 fb GeV$^{-1}$ $\times$ $5.10^{-4}$ = 0.05 fb.GeV$^{-1}$ or 0.2 events in 15 fb$^{-1}$ in a 250 MeV bin. Note however that for the exclusive Higgs production process $n_{ass} =$ 0, while for generic $DY/Z$ production $< n_{ass} > \approx 16$ [26] for $p_T \geq 0.2$ GeV, $|\eta| \leq 1$. The observation of lepton pairs with ***no*** associated tracks, $n_{ass} = 0$, would already be good evidence for exclusive Higgs production [6]. The $CH$ cross section $\sigma(p\bar{p} \rightarrow pH\bar{p})$ $\times$ branching fraction $H \rightarrow \tau^+\tau^-$ of 3.4 (1.3) fb at 115 (130) GeV gives 21 (10) events on a background of less than 1 event if we include a 50% acceptance/efficiency factor. High $p_T$ $\tau$ are easily recognized: one-prong decays are 85% and three-prong are 15%. A high $p_T$ 3-prong $\tau$ decay is quite distinct from a QCD hadronic jet because it is tightly collimated, with $M_{eff} < M_\tau = 1.78$ GeV. From the two neutrinos we will have central mass $M_X < MM$. For any non-diffractive background we can assume that the associated charged multiplicity on the $l^+l^-$ vertex is Poisson-distributed with a mean of about 16, which is what CDF observes [26] for $Z$ events [7]. This non-diffractive background then has a completely negligible tail at $n_{ass} = 0$. Thus the backgrounds in all the dilepton channels with $n_{ass} = 0$ are negligible, and even 3 or 4 events at the same $MM$ would constitute a discovery. Although we only considered fully exclusive production in the above discussion, sometimes the $H$ will be accompanied by some hadrons, central enough to have well measured four-momenta $p_5 ... p_k$. Then one can still use $MM^2 = (p_{b1} + p_{b2} - p_3 - ... - p_k)^2$. Being able to use these "nearly exclusive" events will increase the rate significantly. This technique can only be used with the leptonic H decays to $\tau^+\tau^-$ and $l^+l^-\nu\bar{\nu}$.

---

[6] One should see a peak at $n_{ass} = 0$ in this multiplicity distribution. This is a study that we are starting now.

[7] We realize that $< n_{ass} >$ will decrease when events have a large $x_F$ $p$ and $\bar{p}$, as the energy available for particle production will be less than normal.

### C. $H \to WW^{(*)}$

The Higgs branching fraction to $WW^{(*)}$ rises from 29% at 130 GeV to 69% (97%) at 150 (170) GeV (see Table 1). Beyond 180 GeV it falls because of competition from the $ZZ^{(*)}$ mode. We consider first the leptonic decay modes of the $W$ because of the spectacular cleanliness of the event vertices: either $ee, e\mu, \mu\mu, e\tau, \mu\tau$ or $\tau\tau$ and no other charged particle tracks ($n_{ass} = 0$), together with large $\not{E}_T$ and the forward $p$ and $\bar{p}$.

Precision timing $\approx 30$ ps on the $p$ and $\bar{p}$ will not only check that they came from the same interaction but can pin down the vertex $z_\circ$ to about 1 cm, to be related to the dilepton vertex known to $\sigma_z \approx 20$ $\mu$m. This cleanliness means that the search is insensitive to the number of collisions in a bunch crossing. Using the missing mass method the Higgs mass can be measured with $\sigma_M \approx 250$ MeV *per event* despite the two undetected neutrinos! To estimate the $WW^{(*)}$ signal we extrapolate the Cudell and Hernandez (1.8 TeV) exclusive cross sections from 150 GeV (11 - 16 fb) to 180 GeV (2.5 - 5 fb). Putting in $BR(H \to WW^{(*)})$, a 10% probability that both $W$ decay leptonically, and assuming that, by using lower than usual trigger thresholds on the central leptons and $\not{E}_T$, we can keep the efficiency at 50%, we find in 15 fb$^{-1}$ 7 events for $M_H = 150$ GeV falling to 2.5 events at $M_H = 180$ GeV. The estimates are for $\sqrt{s} = 1.8$ TeV; for $\sqrt{s} = 1.96$ TeV the production cross section will be ($\approx 25\%$) higher. To estimate the background we refer to the observation of five $W^+W^-$ events by CDF [27] [8] which gave $\sigma(p\bar{p} \to W^+W^-X) = 10.2 \pm 6.5$ pb which we assume to be roughly uniform over $160 < M_{WW} < 180$ GeV so $d\sigma/dM \approx 0.5$ pb GeV$^{-1}$. Below

[8] DØ earlier found one event [28] in 14 pb$^{-1}$.

160 GeV the cross section for $WW^*$ will be smaller. The observed $W^+W^-$ cross sections are consistent with Standard Model NLO expectations, ignoring the Higgs, of $\sigma(p\bar{p} \to W^+W^-X) = 10$ pb at 1.8 TeV. We multiply by the 10% probability that *both* $W$ decay leptonicaly and apply a 50% "efficiency" for detecting the $p, \bar{p}$ and both leptons and recognizing the event as $l^+l^-\not{E}_T$. This is high compared with the efficiency in ref [3], which was 5.4% - 8.9%, because due to the lack of background we can lower the selection cuts on $\not{E}_T, p_T(e), p_T(\mu)$ and $p_T(\tau)$ significantly. We assume that about $5 \times 10^{-4}$ of these are from $DI\!PE$, giving $\approx 3 \times 10^{-3}$ fb/250 MeV.

It is also possible to use the exclusive $W^+W^-$ events where one $W$ has decayed hadronically giving $l\nu jj$ with $M_{jj} \approx M_W$. Improvements to our jet algorithms will be valuable. The number of these events is a factor 4.3 more than the number of $l^+l^-\nu\bar{\nu}$ events, and the continuum background is still very small. With the $CH$ estimate this gives 11 events even for $M_H = 180$ GeV, at $\sqrt{s} = 1.8$ TeV, as shown in Table I.

In order not to be limited by the number of interactions in a bunch crossing we will not use a method requiring rapidity gaps (as normally measured in counters or calorimeters). This is where the strength of using only leptonic decays of the $W^+W^-$ enters. Tracking back the $l^+$ and $l^-$ to their common vertex (which can be done using the SVX detectors to a precision $\sigma_x = \sigma_y \approx 10$ $\mu$m and $\sigma_z \approx 20$ $\mu$m ) there will, for the exclusive process, be no other particles coming from the same vertex, $n_{ass}$ $= 0$. All "normal" production of $W$-pairs will on the contrary have a highly active vertex with many associated hadrons. One can plot the missing mass $MM$ for the superclean events with two and only two oppositely charged leptons on a vertex, with and without $\not{E}_T$. A Higgs signal will be a cluster of events at the same $MM$ within the resolution. However as stated above we will also use events with $n_{ass} \neq 0$ and search in $WW \to l\nu jj$ events.

If the exclusive cross section is indeed big enough to provide events in the data, but continuum background were to be an issue, one has further recourse to angular distributions [29]. The $H$ is a scalar and decays isotropically, while generic $W^+W^-$ production is not isotropic with respect to the beam axis. Also the $W$'s (like the $\tau$'s) from a Higgs must have opposite polarizations. This is not generally true for the backgrounds, so one can plot quantities sensitive to these kinematic features as a function of $MM$ to look for localized structure. No matter what the decay mode(s), we can measure $M_H$ with $\sigma_M \approx \frac{250 MeV}{\sqrt{n}}$ (statistical). The systematic uncertainty will be determined by how well we can calibrate the $MM$ scale using elastic (section IIIA) and low mass exclusive inelastic events (section IIIB).

### D. $H \to ZZ$

Looking for $H \to ZZ$ at the highest masses 190(200) GeV where its SM branching fraction is 0.219(0.261) presents special challenges. The $(CH)$ exclusive production cross section is about 1.5(0.5) fb (based on an extrapolation), about 1%(0.3%) of the inclusive $gg \to H$ production. Putting in the branching fraction to $ZZ$ we find only about 5(2) events in 15 fb$^{-1}$. On the other hand we can perhaps use *all* decays including $\nu\bar{\nu}\nu\bar{\nu}$ ($MM \approx 200$ GeV with *nothing* on the primary vertex! But this is only 4% of the decays.) We can perhaps pick up some cross section by allowing a few measured hadrons on the primary vertex. The cases where one $Z$ decays to $\nu\bar{\nu}$ and the other decays visibly (32%) are interesting in that the invisible missing mass $(p_1 + p_2 - p_3 - p_4 - p_{Z-visible})$ should be equal to $M_Z$. In the 42% of the cases where only one $Z$ decays to jets we can apply the $M_{JJ} = M_Z$ constraint. While the $ZZ$ channel is very marginal with 15 fb$^{-1}$ it could become interesting with higher luminosity.

## IV. EXCLUSIVE $\gamma\gamma$ PRODUCTION

Fortunately there is a process that is very closely related to exclusive Higgs production, namely the exclusive production of two photons by $gg$-fusion through a quark loop. While in the Higgs case only the top quark loop is significant, in this case all quarks contribute, although the up-type quarks contribute a factor $Q^4 = 16$ more than the down-type quarks. The crucial similiarity is that in both cases the final state, $H$ or $\gamma\gamma$, is not strongly interacting[9]. Therefore the non-perturbative parts of the process should be *identical* in exclusive $\gamma\gamma$ and $H$ production. The ratio

$$\frac{d\sigma(M)}{dM_{\gamma\gamma}} : \sigma_H(M)$$

should be theoretically well predicted (although we cannot measure both at the same $Q^2$), and related to the inclusive ratio (selecting the $gg$ part of the $\gamma\gamma$ production). A calculation including helicity effects has not yet been done. We can measure $p\bar{p} \rightarrow (p)\gamma\gamma(\bar{p})$ as a function of $M(\gamma\gamma)$ and that should give us a reliable estimate of $p\bar{p} \rightarrow pH\bar{p}$. In 2 fb$^{-1}$, if the exclusive fraction is $10^{-3}$ we will find 13 exclusive events in the mass bin 10 - 40 GeV(we have applied a reduction factor of 0.093 to have CDF otherwise empty, at L = $10^{32}$ cm$^{-2}$ s$^{-1}$). If the $\gamma\gamma/H$ ratio can be reliably predicted, even if we do not *find* the Higgs we might be able to exclude it over some mass range. HERWIG [30] calculations of the $\gamma\gamma$ production in 15 fb$^{-1}$ are given in Table II, together with the numbers of exclusive events we would find if the exclusive fraction is $10^{-3}$ of $gg \rightarrow \gamma\gamma$. This study will be done without attempting to detect the $p$ and $\bar{p}$, so all $t$ and $\phi$ values are accepted. We are not likely to find any exclusive $\gamma\gamma$ events with the $p$ and $\bar{p}$ detected.

We are able to start such a study now, without seeing the $p$ and $\bar{p}$ but looking for events that have two photons, fairly well balanced in $p_T$, and nothing else visible in all the CDF detectors, including the forward Miniplugs and Beam Shower Counters. To do this we will include a trigger on two electromagnetic towers with $E_T > 5$ GeV (3 GeV if possible) with a Level 1 veto on the Miniplugs and BSC. At Level 2 (or 3) we require zero tracks and no energy in the hadronic calorimeters. These requirements will veto crossings with any additional inelastic interaction, so the useful luminosity is reduced by a factor $e^{-<n>}$ where $<n>= L\sigma_{inel}\Delta t$, $\sigma_{inel} = 60$ mb and $\Delta t = 396$ ns so at $L = 1.0\ \times 10^{32}$ cm$^{-2}$s$^{-1}$ we have $<n> = 2.4$ and $e^{-<n>} = 9\%$. (When we see the $p$ and $\bar{p}$ we will not have to apply this factor.)

---

[9]The $b$ and $\bar{b}$ from $H$ decay do not count, as the "light" Higgs is a stable particle ($\Gamma < 10$ MeV) on the strong interaction time scale. For the heavy Higgs we can just look at leptonic decays of the $WW^{(*)}$.

| $M_{\gamma\gamma}$(GeV) | $N_{all}$ | $N_{gg}$ | $10^{-3}N_{gg}$ |
|---|---|---|---|
| 10-20 | 823814 | 822992 | 823 |
| 20-40 | 297905 | 227866 | 228 |
| 40-60 | 44271 | 21930 | 22 |
| 60-80 | 14117 | 4591 | 4.6 |
| 80-100 | 4454 | 1439 | 1.4 |

TABLE II. The numbers of events expected in HERWIG in 15 $\text{fb}^{-1}$ with $|\eta_\gamma| < 2$ and $p_T(\gamma) > 5$ GeV/c. The third column shows the number produced in $gg$ collisions and the last column the number of exclusive photon pairs if the fraction is $10^{-3}$ of $N_{gg}$.

We have inclusive $\gamma\gamma$ data from Run 1 and are starting to look for evidence of single diffractive or double pomeron rapidity gap signals. However this is just a "warm up" exercise as we do not expect more than $10^{-2}$ (and it could be much less) of those events that come from $gg$ fusion (not $q\bar{q}$ annihilation) to be exclusive.

## V. BEYOND THE STANDARD MODEL

### A. Extra generations

If there exist more massive strongly interacting fermions than the top quark the $gg \rightarrow H$ cross section will be enhanced as the additional loops come into play. One intriguing possibility, that there exists a 4th generation of very massive and nearly degenerate quarks and leptons, has been recently proposed by S. Sultansoy [31]. The LEP generation-counting experiment would not have been sensitive to this because the "neutrino" is too massive. Sultansoy's expectation is that $m_4 \approx 8m_W$, in which case the Higgs production cross section is enhanced by a factor of approximately 8. (This affects the $gg$-fusion process of interest to this letter of intent, but not the more orthodox $W^* \rightarrow WH$ associated production process.)

### B. Extended Higgs models, CP-odd scalars

In extended Higgs models [32], a Higgs boson $h^\circ$ may have quite different decay modes from the SM modes considered in the previous section. One possibility is for the $h^\circ$ to decay to a pair of light neutral CP-odd scalars $A^\circ$ which have a supressed coupling to fermions. This could be the dominant decay mode, rather than the $b\bar{b}$ mode for the lighter Higgs masses. The $A^\circ$ may be even lighter than 0.5 GeV and will then decay with nearly 100% branching ratio to $\gamma\gamma$ which would not be resolved, so the event would look like two high-$E_T$ direct photons. Higher mass $A^\circ$ can give similar signatures, e.g. $h^\circ \to A^\circ A^\circ \to 3\pi^\circ + 3\pi^\circ$. Even though there may be no associated hadrons on the primary vertex we can still find $z_o$ from the timing on the $p$ and $\bar{p}$.

### C. Top-Higgs

Another example of accessible Higgs physics beyond the standard model is the idea of the top-Higgs, $h_t$, in which a $< t\bar{t} >$ condensate is responsible for the large top quark mass. In the Topcolor Assisted Technicolor (TATC) scenario proposed by Hill [33], the top-Higgs is a $t\bar{t}$ bound state which could be as light as 200 GeV. As $t\bar{t}$ decays are kinematically forbidden the predominant decay is to $t\bar{c}$ or $\bar{t}c$ [34]. The production cross section via gluon-gluon fusion could be nearly 1 pb at the Tevatron with $\sqrt{s} = 2$ TeV. If $10^{-3}$ of these $h_t$ are produced exclusively then 15 fb$^{-1}$ would produce 15 events of the type $p\bar{p} \to p + t\bar{c} + \bar{p}$. The $h_t$ width is expected to be $< 7$ GeV.

### D. Lightest SUSY particle

There are light mass windows where a $\tilde{\chi}_1^\circ$ is not excluded (see e.g. ref [35]). If R-parity is conserved then $\tilde{\chi}_1^\circ$ would be long-lived or stable and only weakly interacting. This state can be produced in pairs, or with a $\tilde{\chi}_2^\circ$, in $gg$ interactions via a $q\tilde{q}$ box diagram. Normally one concentrates on the $\tilde{\chi}_1^\circ\tilde{\chi}_2^\circ$ associated production, because the (much more massive) $\tilde{\chi}_2^\circ$ can be detected through its decays, while $p\bar{p} \to \tilde{\chi}_1^\circ\tilde{\chi}_1^\circ + X$ does not have a distinctive final state, both $\tilde{\chi}_1^\circ$ being invisible. However our $MM$ technique provides a possibility. We select events with a measured $p$ and $\bar{p}$ and plot the $MM$ spectrum for events where there are no particles on the primary vertex ($z_\circ$ coming from the timing). We exclude elastic scattering by cutting on $\Delta t$ and $\Delta\phi$. We have to exclude additional interactions by requiring no tracks, and only noise in all the CDF calorimeters (including the $BSC$s). We then search the $MM$ spectrum for a localized threshold effect (a step). The main background is from interactions like $p\bar{p} \to p\pi^\circ\pi^\circ\bar{p}$ or $pn\bar{n}\bar{p}$ or $pK_L^\circ K_L^\circ\bar{p}$, where the central hadrons are too soft to distinguish from noise, or which go in detector cracks. The effective luminosity for this search will be lower because of the single interaction requirement. The *optimum* luminosity for this is about 1.2 $10^{32}$ cm$^{-2}$ s$^{-1}$ when the average number of inelastic interactions per crossing is 1.0 (when $\Delta$t = 132 ns).

### E. Color Sextet Quarks

It is conceivable that the dynamical "Higgs mechanism" that gives the $W$ and $Z$ bosons their mass involves Goldstone bosons composed of color sextet quarks [36] $Q_6$. Massive color sextet quarks may also exist without providing the Higgs mechanism. In the latter case the the additional $Q_6$ loops in the process $gg \rightarrow H$ will substantially increase the Higgs production cross section. However a $H$ is not needed since the electroweak symmetry breaking can be generated by chiral symmetry breaking in the $Q_6$ sector. Pomeron-pomeron interactions would be the ideal place to expose this physics. $W$-pairs will be produced with a relatively large cross section once $\sqrt{\hat{s}_{I\!P I\!P}}$ exceeds $2M_W$. This is rather marginal for the Tevatron but will not be for the LHC, and we should certainly look. There can be other manifestations, such as the $\eta_6$, which is like the Higgs in many respects but will be produced with a much larger cross section (a normal strong interaction cross section at high enough energies).

### F. Graviton emission

If there exist "large" extra dimensions in which (Kaluza-Klein) gravitons can propagate, while the known particles are confined to the 3-dimensional "3-brane", one can explain the relative weakness of gravity. Gravitons $\mathcal{G}$ can be created in $p\bar{p}$ collisions either singly through $gg \rightarrow \mathcal{G}$ (with or ... in our case ... without a recoiling gluon) or with much lower cross section in pairs $gg \rightarrow \mathcal{G}\mathcal{G}$. Gravitons probably exist as a large (or infinite) number of states of different mass: a "graviton tower". Gravitons emitted into "the bulk", out of our 3-D world, will be invisible except in so far as they will give rise to an *apparent* violation of 4-momentum conservation. This is how neutrinos were first "seen". Our proposed experiment is ideally suited to search for the emission of such states [37]. We select non-elastic events where there are no central tracks and all the CDF detectors (except the pots) are consistent with being empty, as in the search for $\tilde{\chi}_1^\circ\tilde{\chi}_1^\circ$. This clearly restricts us to single interactions/crossing, but even at $L = 2.10^{32}$ cm$^{-2}$ s$^{-1}$ with 132 ns between crossings we have $\approx 1.5 \times 10^6$ s$^{-1}$. The $MM$ spectrum might have five components:

(1) The $KK$ graviton tower production signature would be a rising then a falling distribution starting at $MM$ = (essentially) 0. Individual levels of the tower will not be resolved. The cross section $\sigma(gg \to \mathcal{G})$ rises with the "effective $M_{\mathcal{G}}$", because the $\mathcal{G}$ couples to the gluons via their stress-energy tensor. The distribution falls after the initial rise because the $gg$ luminosity falls with increasing $\sqrt{\hat{s}(gg)}$. The cross section can be calculated [38] and (extrapolating down in $E_T$) is expected to be about 10 pb, giving 10,000 events in 1 fb$^{-1}$. Only a fraction ($\approx$ 1% ?) of these will be exclusive. We can also use the "nearly exclusive" events, using the equation

$$MM^2 = (p_{b1} + p_{b2} - p_3 - p_4 - \sum_{i=5}^{n} p_i)^2$$

where the sum includes all the particles measured in the CDF central detectors.

(2) Lightest SUSY particle ($\tilde{\chi}_1^\circ$) pair production, giving a threshold rise at 2 $M_{\tilde{\chi}_1^\circ}$. We would not complain if this was our background!

(3) Elastic scattering where the $p$ and $\bar{p}$ came from different events, their partner $\bar{p}$ and $p$ having been missed. This will have to be Monte-Carlo'd but with 360° forward track coverage it should be very small, and will give $MM \approx 0$ (or somewhat negative) as both $p$ and $\bar{p}$ have the beam momentum but are not colinear.

(4) Low mass $DI\!PE$ where the central state is missed, perhaps being $n\bar{n}$ or $K_L^\circ K_L^\circ$, although these should be detected in the calorimeters, especially the $\bar{n}$ which deposits a 2 GeV annihilation signal. Unfortunately there are calorimeter cracks so a purely neutral final state can fake an empty event. This will need to be studied by simulations.

(5) Double beam halo events. We can measure this by combining $p_{halo}$ tracks with $\bar{p}_{halo}$ tracks. $10^n$ events of each type give $10^{2n}$ $p_{halo}\bar{p}_{halo}$ combinations. Even if the beam halo conditions fluctuate from run to run, we can use halo tracks found in coincidence with elastic scattering events which have no coincident inelastic events. Thus the halo-halo fakes are monitored continuously. One can even do this bunch-by-bunch (and the information might be useful for Tevatron diagnostics).

Tachyons $\mathcal{T}$ are hypothetical states which *always* have speeds in excess of $c$,

so $|p| > E$ and $M_T^2$ is negative. Zero energy tachyons have infinite speed. As $E_T$ increases the speed tends to $c$ *from above*. If there are extra *time* dimensions then KK towers of tachyons could exist [39], giving rise to a continuum in the $MM^2$ spectrum for negative $MM^2$ ! To us confined to a 3-brane we see the $p$ and $\bar{p}$ coming out of the collision with more energy than they had initially! This is bound to be more interesting than cold fusion!

### G. Micro Black Holes, MBH

If there are large extra dimensions and gravity is strong in the (say) 10-dimensional world, then micro black holes MBH should exist on a mass scale $M^*$ corresponding to the size of the extra dimensions. This is not likely to be $< 500$ GeV and so this physics (like all the physics of this section) is probably not accessible with leading $p$ and $\bar{p}$ at the Tevatron. We are including it here because this may well be the best way to study MBH in hadron-hadron collisions at the higher energies of the LHC and VLHC, and we can start to learn about it at the Tevatron. Also, this experiment is exploratory and, *you never know!* Even if $M^*$ is in the TeV range MBH in the hundred-GeV range will still exist (in the decay of a TeV MBH particles are emitted and lower mass MBH's are created). The production of a (say) 200 GeV MBH will proceed at a rate less than the strong interaction rate but it will still occur.

In collaboration with Liubo Borissov and Joe Lykken [40] we are studying MBH production by $gg$ fusion and decay in $pp$ and $p\bar{p}$ collisions. Once the $gg$ energy reaches the scale $M^*$ this is expected to proceed with a cross section typical of the strong interaction. For example if $M^*$ *were* to be as low as 200 GeV (which is probably excluded because the di-jet mass spectrum is well fit by QCD out to higher masses) then $\sigma(p\bar{p} \rightarrow MBH + X) \approx 100$ nb. Suppose we put in a factor $\frac{1}{64}$ to require the MBH to be in a color singlet, and another factor $10^{-4}$ ("educated guess") to require two large rapidity gaps with a leading $p$ and $\bar{p}$, we get $\sigma(p\bar{p} \rightarrow p + \text{MBH} + \bar{p}) \approx 150$ fb. We can allow additional low $p_T$ hadrons to be produced along with the MBH, as we are not looking for a narrow state. We just need to measure the $p$ and $\bar{p}$ in order to measure the total mass of the central system, MBH + hadrons.

What happens to the MBH once they are produced? They decay promptly[10] (on a strong interaction time scale) to anything that couples to gravity, i.e. *anything*. They decay into photon pairs, neutrino pairs, gravitons, $e^{\pm}, \mu^{\pm}, \tau^{\pm}$, quark pairs, $W$ and $Z$ pairs if massive enough, etc. These MBH are very hot and tend to decay to a few very energetic particle pairs. We show in fig 8 the first twenty 200 GeV MBH generated

| | | | |
|---|---|---|---|
| GG 138.00<br>ss 36.64<br>cc 18.95<br>$d\bar{d}$ 5.03<br>ss 1.18 | $\gamma\gamma$ 121.60<br>$\nu_\tau\bar{\nu}_\tau$ 34.44<br>$d\bar{d}$ 43.26<br>$e^+e^-$ 0.57 | $\nu_\tau\bar{\nu}_\tau$ 180.50<br>$\tau^+\tau^-$ 19.11<br>$\gamma\gamma$ 0.25 | cc 183.60<br>GG 13.85 |
| GG 164.90<br>GG 28.11<br>$d\bar{d}$ 2.77<br>$\nu_\mu\bar{\nu}_\mu$ 3.56<br>$\gamma\gamma$ 0.33 | GG 176.20<br>$\tau^+\tau^-$ 21.46<br>$e^+e^-$ 1.99 | ss 138.70<br>$\gamma\gamma$ 45.61<br>$\tau^+\tau^-$ 13.39<br>$u\bar{u}$ 1.50<br>$u\bar{u}$ 0.62 | $\tau^+\tau^-$ 83.80<br>$b\bar{b}$ 56.74<br>ss 40.35<br>$e^+e^-$ 17.00<br>$u\bar{u}$ 1.68 |
| $\nu_e\bar{\nu}_e$ 198.30 | $W^+W^-$ 165.90<br>$u\bar{u}$ 28.98<br>cc 4.85 | $\gamma\gamma$ 125.30<br>$b\bar{b}$ 31.25<br>$\gamma\gamma$ 40.44 | ZZ 184.00<br>$\gamma\gamma$ 14.55<br>$\nu_e\bar{\nu}_e$ 1.07 |
| GG 194.20<br>$d\bar{d}$ 5.65 | $\gamma\gamma$ 151.40<br>$\nu_\tau\bar{\nu}_\tau$ 38.26<br>$e^+e^-$ 3.29<br>$\mu^+\mu^-$ 4.95<br>$e^+e^-$ 1.56<br>$d\bar{d}$ 0.32 | cc 85.61<br>GG 111.60<br>$\nu_\mu\bar{\nu}_\mu$ 1.83<br>$\gamma\gamma$ 0.53 | $\gamma\gamma$ 145.40<br>$d\bar{d}$ 37.46<br>$u\bar{u}$ 12.98<br>$\gamma\gamma$ 2.38<br>$e^+e^-$ 1.58 |
| GG 167.10<br>GG 18.38<br>GG 11.51<br>$\gamma\gamma$ 1.36 | $b\bar{b}$ 196.20<br>$\nu_\mu\bar{\nu}_\mu$ 3.19 | GG 124.10<br>GG 64.06<br>$u\bar{u}$ 9.76<br>$\nu_e\bar{\nu}_e$ 1.45 | $\gamma\gamma$ 155.60<br>GG 43.58 |

FIG. 8. The first 20 Micro Black Hole events generated, fixing their mass to be 200 GeV. All particles with $E > 100$ MeV are shown.

Each member of the pair carries the same energy (we are in the c.m. of the MBH). Seven of the 20 events have more than 100 GeV/200 GeV taken away by gravitons, and two have more than 180 GeV in neutrinos. Five events have a $\gamma\gamma$ with $M_{\gamma\gamma} > 100$ GeV. One event has a $W^+W^-$ and one has a $Z^\circ Z^\circ$. These are very striking events, which become even more dramatic as $M^*$ increases. Independent of the VFTD we should certainly search for events of this type in Run 2[11]. However the events where nearly all the energy is taken away in gravitons or neutrinos will be invisible unless the forward $p$ and $\bar{p}$ are detected.

---

[10]Therefore they do not have time to eat other particles and grow!

[11]Perhaps the mysterious $ee\gamma\gamma\not{E}_T$ event in CDF Run 1 [41] is a MBH!

Although a proper study has not yet been done, the scale $M^*$ is probably $>$ 800 GeV because the dijet mass spectrum [42] is well fit by a QCD calculation up to 1 TeV. (If 800 GeV MBH could be formed with the strong interaction cross section they would reduce the jet yield in favor of the more exotic final states.)

We will of course look for exotic final states in this proposed experiment and this will at least be valuable experience for later searches at LHC and VLHC.

## VI. TRIGGERS

This is a very rich physics program and we would like a powerful trigger system to make optimum use of the luminosity with minimum impact on the rest of the CDF program. To do this we propose a fast trigger processor working at Level 2.

Prompt signals come from the solid Cerenkov counters and from the front scintillator, and an "arm" trigger will be a coincidence between these. We have four arms : NE, SE, NW, SW. The 2-arm trigger will be based on (NE + SE) * (NW + SW) in coincidence with the beam crossing signal, $X$.

The next stage of the trigger, at Level 2, is to look for and compute the tracks : $x_1, y_1, \frac{dx}{dz}, \frac{dy}{dz}$. The trigger processor for this will be based on the existing SVT, which finds tracks at Level 2 using hits in the central silicon tracker SVX. A trigger processor will calculate the missing mass $MM$ using $x_1, y_1, \frac{dx}{dz}, \frac{dy}{dz}$ on each arm and the vertex $z_o$ from $(t_E - t_W)$. Different ranges of $MM$ will be separately prescalable, and put in combination with other requirements at Levels 2 and 3. Central requirements will be a combination of jets (including hadronic $\tau$ decay), $e$'s and $\mu$'s, $\gamma$'s, $\not{E}_T$, and also *nothing* $\emptyset$ visible on the interaction point (from the FTC). In the latter case elastic scattering will be separated out using $\Delta t$ and $\Delta\phi$ cuts. Elastic scattering events will be recorded without the main CDF detectors, and probably with a veto on the TOF, BSC, Miniplugs, and perhaps more. So the events will be very small and there may be no reason not to take the full rate of $\approx 10$ s$^{-1}$. The reason for wanting to record elastic events where CDF is empty, therefore without a coincident inelastic interaction, is to build up a library of beam halo tracks. These show up as random coincident tracks in the pots, and (a) are important for reconstructing the halo-halo background (b) can be used, bunch-by-bunch, as a machine diagnostic of halo(x,y). Inelastic events $p\bar{p} \rightarrow p\emptyset\ \bar{p}$ will be especially interesting, in combination with no significant signals in all the calorimetry, the TOF, the CLC and BSC. We do not yet have rate estimates for all the various triggers that we propose. We can live within a specified bandwidth by prescaling, but do not want (and will not need) to compromise on the 115 - 180 GeV Higgs search. We suggest that an incremental data rate (to tape) up to 5 $s^{-1}$ should be allowed for this program. (Incremental, because some of the events will be triggered on anyway, and we will read out our detectors for *every* CDF event.) Note that the elastic and low mass events will be very small.

## VII. TIMESCALE

At present, assuming a $2-3$ month shut-down starting in late 2002, we foresee the following schedule:

• March 2001: We ask the Director to transmit this proposal to the PAC them to take note of it at the April 20th Meeting. We request that the Directorate ask the Beams Division to evaluate the consequences of the proposed Tevatron modifications and provide a cost and statement of the time needed. We request that the Directorate ask CDF to evaluate the effect on its baseline program, and to judge the detectors and integration of the DAQ and trigger. We ask the Director to allow us to present the proposal, updated with costs and firmer timescale, to the June PAC so that a decision can be taken after a recommendation from the November PAC meeting.

• Jan - Oct 2001: Technical design of detectors and vacuum vessels. More detailed tracking calculations. Monte Carlo study of acceptances with different final states. Design of trigger processor.

• November 2001: Final approval by PAC.

• 2001 - fall 2002 Construction of detectors and all hardware (including mechanics for vacuum system, electronics for DAQ and trigger).

• Summer 2002: Beam tests of "9th pot" with detectors.

• Fall 2002 or as soon as 2-month shutdown occurs: Modifications to the Tevatron and installation of roman pots and any detectors which are ready.

• As soon as possible thereafter: Installation of all detectors, beam commissioning, trigger tests, and data.

## VIII. COSTS

We cannot give reliable cost estimates at this stage because the cost of the Tevatron modifications and the pots and vacuum chambers have not yet been established. However the best estimate we can make at this time is less than $ 0.8 M. More information Will be available for the June PAC meeting, when we plan to submit a full proposal.

• Modifications to the Tevatron : see section IC. Costs will only come after a Beams Division study.

• Silicon detectors FST: Four telescopes of eight planes each, + four spare planes in test beam pot. Number of channels 2048 per plane, 73,728 total. 36 Sensors @ 2.8 K$ = 100 K$, hybrids, SVX4 chips, DAQ, power supplies $245K ×1.5(contingency) = $368K.

• Four (+1) half disc trigger scintillation counters with twisted strip light guides and R5900U PMTs, read-out and DAQ (5 channels).

• Four (+1) Time of Flight Cerenkov counters FTC, each one being a hodoscope of 5 counters with a R5900U PMT, plus read-out and DAQ (25 channels)

• Cables from detector stations to BØ .

• Trigger electronics, with special Level 2 $MM$ processor. This will be studied before the June PAC.

• Detector stations, including motors, position sensors, slow controls, precision position read out. These will be built at Helsinki.

## IX. PEOPLE

This will, if approved, become an integral part of CDF like other "Beyond the Baseline" proposals such as the central TOF system and Layer 00. We are however bringing additional people and resources to CDF for this project. The Helsinki

Group, from the University of Helsinki and the Helsinki Institute of Physics, comprises R.Orava (group leader), R.Lauhakangas, S.Tapprogge and K.Österberg. The Helsinki group are not yet members of CDF but they are applying to join to participate in this VFTD project. They will take the responsibility of constructing the vacuum chambers with the roman pots, in close consultation with Fermilab staff. They will also produce the hybrids for the FST. S.Tapprogge will also work on the triggers and R.Lauhakangas on the DAQ. Some other Fermilab staff will participate in this proposal without becoming full members of CDF; they will sign all papers resulting from this proposal. These are D.Finley (Technical Division) and C.Moore (Beams Division) who will work on all aspects to do with the Tevatron. M.Albrow, M.Atac (both Fermilab) and A.Rostovtsev (ITEP, Moscow) will be responsible for the trigger counters (scintillators and FTC). P.Booth and S.Marti-Garcia (Liverpool) and others in the Liverpool group will be responsible for providing the sensors for the FST. M.Lancaster and other UCL,London people (P.Crosby, D.McGivern, A.Wyatt and R.Snihur) will work closely with the Liverpool group. B.Heinemann (Liverpool) and R.Snihur (UCL) are working on tracking reconstruction software and Monte Carlo generators. Liverpool have a processor farm [43] which has the capacity to do all the Monte Carlo and reconstruction of the data. D.Litvintsev (Computing Division) will also work on these aspects and make the event displays. T.Liu (Wilson Fellow) will work on the trigger. W.Wester (Fermilab) will work on all detectors and on triggers. M.Albrow will be the contact person for the VFTD. We expect more students and post-docs from Liverpool and Helsinki to join. While we already have a strong enough team to carry out this project, we welcome additional people.

## X. ACKNOWLEDGEMENTS

This work was supported by the U.S. Department of Energy, the U.K. Particle Physics and Astronomy Research Council (PPARC) and the Institute for Theoretical and Experimental Physics (ITEP),Russia. We thank D.Augustine, P.Bagley, M.Church J.Johnstone and J.Theilacker for information on the Tevatron, and P.V. Landshoff, V.Kim, D.Kharzeev, E.Levin, V.Khoze, A.Martin and M.Ryskin for discussions on exclusive Higgs production. We acknowledge helpful exchanges with J.

Hewett and J. Lykken about extra dimensions, and with L.Borissov and J.Lykken about Micro Black Holes. We acknowledge useful interactions with T.Bowcock and R. McNulty about the silicon detectors.

## XI. APPENDIX I: FORWARD SILICON TRACKER, FST

The silicon microstrip detectors will be the same as those being developed at Liverpool University for the VELO (VErtex LOcator) detector of LHCb. Spatial resolution is an important issue. Radiation hardness is much less critical for the FST than for the centrally located (in high luminosity) VELO detectors. It is possible to design silicon sensors that operate after radiation doses of $10^{15}$ p/cm$^2$ ($\approx$ 40 Mrad).

The design of the VELO silicon sensors of LHCb is not final, but prototyping and testing is underway. Silicon sensors were manufactured by MICRON Semiconductor. They consist of single sided $p^+n$ detectors ($p$-type strips in a $n$-type bulk). The bulk was oxygenated in order to improve the radiation tolerance [45].

The detectors are half discs with a semicircular cut-out for the beams. The $r$-detectors have circular strips, and the $\phi$-detectors measure the azimuth. The latter strips are skewed by a few degrees (depending on $r$) *wrt* a radial vector, and one of the two in a roman pot will be reversed to give stereo information. Figure 9 presents a schematic view of the $r$- and $\phi$- detectors and their strip design.

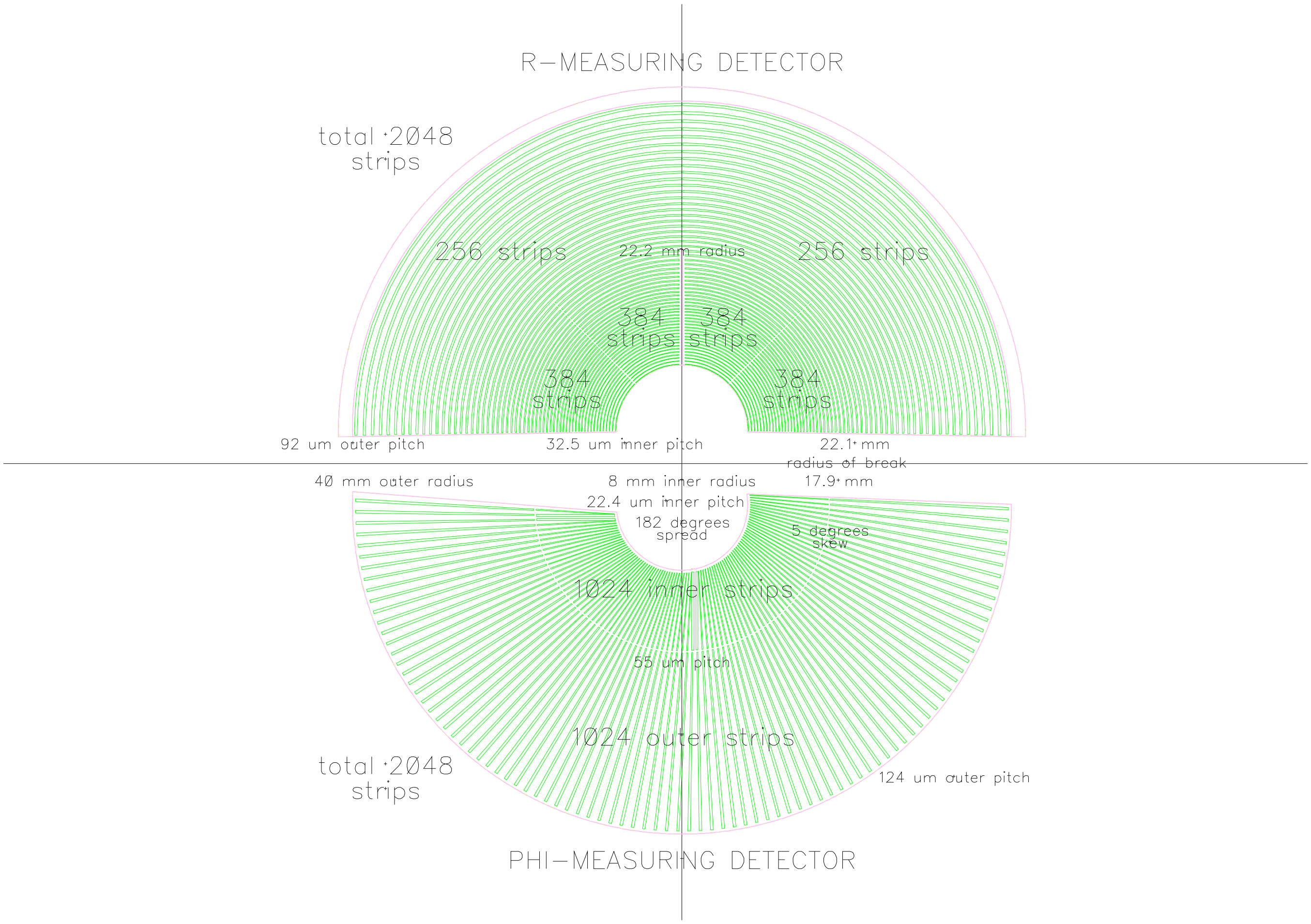

FIG. 9. Schematic view of the strip geometry for an $r$-detector (top) and a $\phi$-detector (bottom). Each has 2048 strips.

The strips are AC coupled to the readout electronics, and there is a double metal layer for readout of the inner strips. Each sensor (both $r$- and $\phi$-detectors) has 2048 strips. Therefore 16 readout chips of 128 channels are needed per detector. We will use 300 $\mu$m sensors.

The $r$-strips are circular arcs. There are four inner segments with 384 strips each and two outer segments with 256 strips each, making 2048 strips total. The four inner segments cover $-91^\circ$ to $+91^\circ$, with gaps between segments of 12, 63, and 12 $\mu$m. There is a 12 $\mu$m gap between the two outer segments. The inner and outer radii are approximately 8 mm and 40 mm. The innermost 189 strips are at constant pitch of 32.5 $\mu$m. The pitch of the outer strips increases in proportion to the radius, to a maximum (strip 640) of 92 $\mu$m.

The $\phi$-strips are straight lines, divided into an inner segment and an outer segment with 1024 strips each. The strips are skewed so that the angles between the strips and a radial line varies with radius from 11.23° at $r = 8$ mm to 2.23° at $r =$ 40 mm.

The estimated cost is approximately 2000 *GBP* ($\approx$ 2.8 K$) per sensor.

Radiation hardness is much less of an issue for the FST than for the LHCb detectors. At $L = 10^{32}$ cm$^{-2}$ s$^{-1}$ we expect $< 10^5$ particles cm$^{-2}$ s$^{-1}$ in the hottest part of the silicon, or $< 10^{13}$ in 5 years of running. It has been proven already that the detectors operate after $10^{15}$ $p$ cm$^{-2}$.

The spatial resolution of the silicon sensors has been measured in a test beam. As the strip pitch is not uniform, the spatial resolution is not uniform across the detector; it is better at small radii. For silicon sensors equipped with analogue readout chips one can get better resolution than the basic strip-pitch/$\sqrt{12}$. The ionization induced charge is shared among 2 or 3 strips. The number of strips in a cluster depends on the projected angle of the particles, and the resolution can be improved by tilting the detectors. Figure 10 shows the spatial resolution attained with 300 $\mu$m thick microstrip detectors from test beam studies [46]. The results presented cover two regions with different strip pitch. Note that the spatial resolution depends on the projected angle of the particles, due to charge sharing among strips. The results presented in figure 10 were obtained with the VA2 analogue readout chip (Viking) [47].

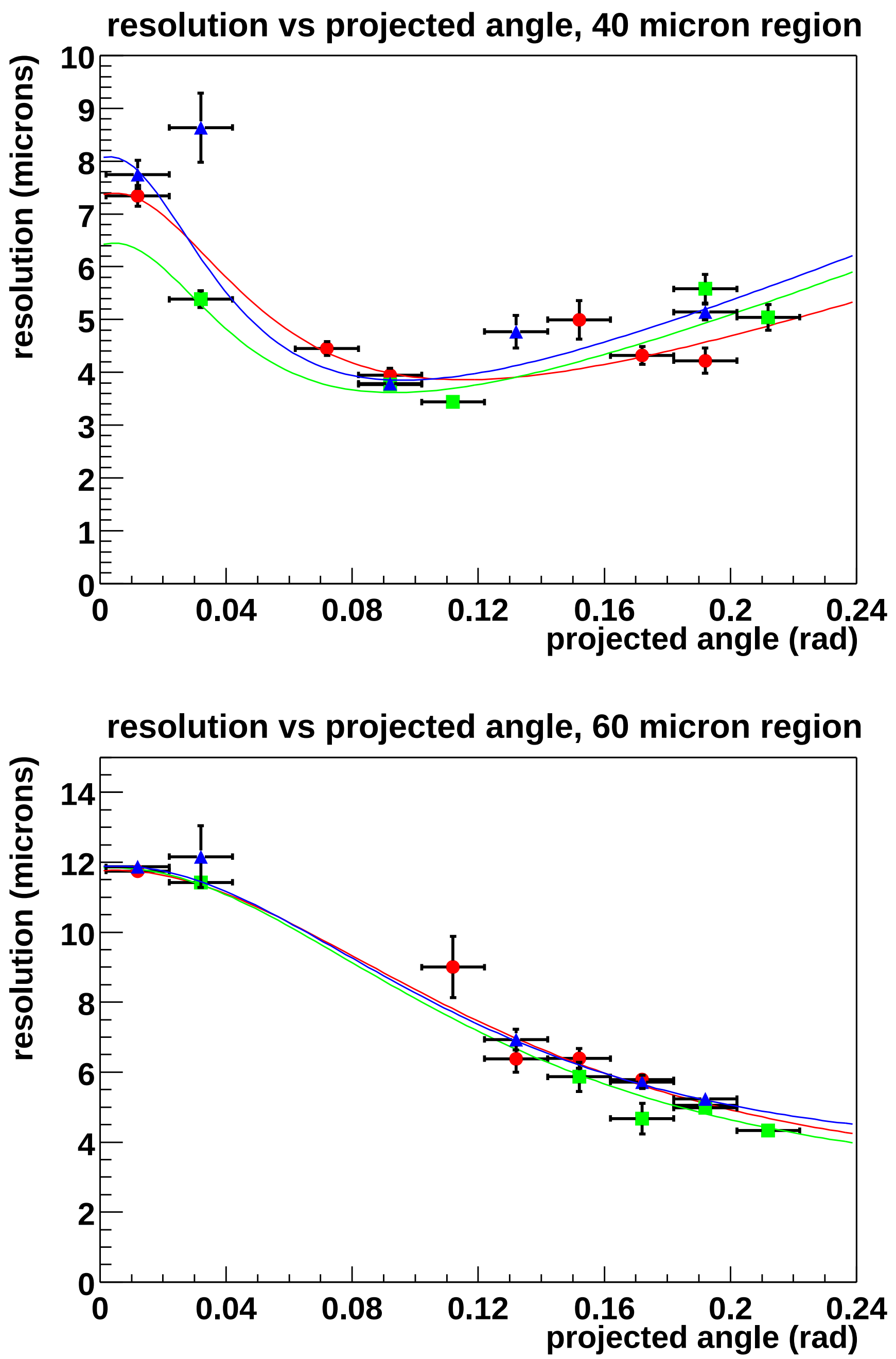

FIG. 10. Resolution from test beam measurements of three 300 $\mu$m detectors with the VA2 readout chip, as a function of angle from the normal. The lines are fits.

## XII. APPENDIX II: STRONG INTERACTION PHYSICS

Although it is often claimed that "we have a theory of strong interactions, namely QCD" that is far from the truth. QCD is a theory of quark and gluon interactions at "large" $Q^2$, and it is quite unable to predict quantitatively *any* interactions of the only things we can detect, namely hadrons. We should not be satisfied that we have a theory of strong interactions until we are able to calculate simple processes such as hadron-hadron elastic scattering. Presumably this future theory will be based on QCD (or QCD will be a high-$Q^2$ limiting case of it), and it will also enable us to calculate Regge-like behavior, as Regge phenomenology provides a rather good decription (certainly the best to date) of simple processes such as

$$\pi^- p \rightarrow \pi^\circ n.$$

We believe that the Tevatron should not be turned off without a measurement of large-$|t|$ elastic scattering, which may provide an important test of such a future theory, and that this proposal is the best that can be done with available technology.

### A. Elastic scattering at high-t.

The detectors will have acceptance for particles with $p = p_{beam}$ $(MM = 0)$ for $|t|$ values of order 0.8 GeV$^2$ - 4.0 GeV$^2$. As the detectors on the $p$ and $\bar{p}$ sides have acceptance for $\Delta\phi = 180°$ we will have good acceptance for elastic scattering events. Elastic scattering in this $|t|$ region has not been measured at Tevatron energies. At the CERN $S\bar{p}pS$ collider, $\sqrt{s} = 540$ GeV, there is structure around 1.0 GeV$^2$ which was seen at the ISR ($\sqrt{s} = 53$ GeV) at higher $|t|$ [48](Fig.11). It would be interesting to measure this at the Tevatron, where we have so far only measured out to -0.6 GeV$^2$ [49].

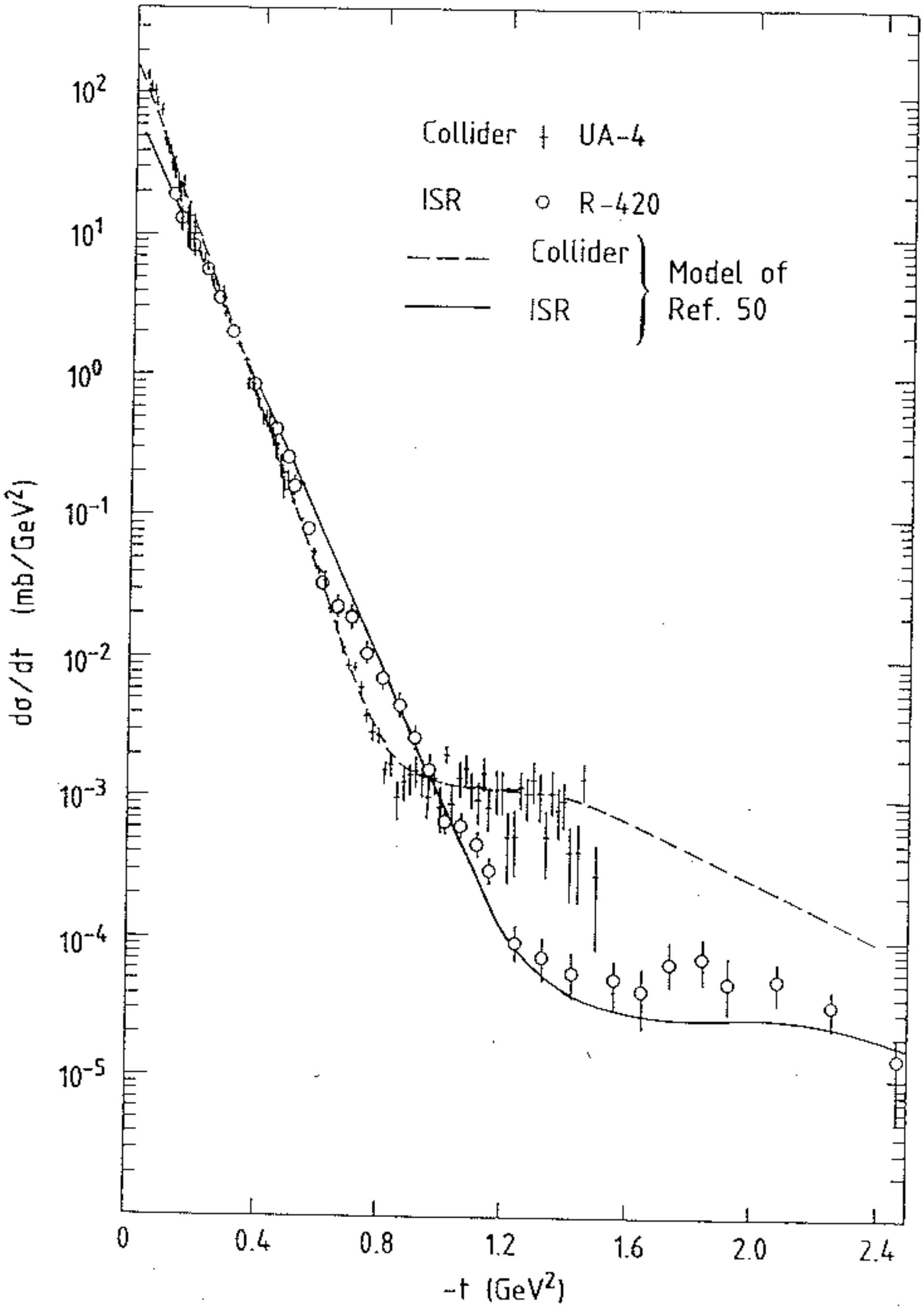

FIG. 11. $p\bar{p}$ elastic scattering at 52.7 and 546 GeV (CERN $Sp\bar{p}S$ Collider).

Donnachie and Landshoff [50] have considered large-$|t|$ elastic scattering in terms of triple gluon exchange (one between each quark pair). Gauron, Nicolescu and Leader [51] have fitted both $pp$ and $p\bar{p}$ elastic scattering and made a prediction for $\sqrt{s} = 1.8$ TeV. In their model large $|t|$ elastic scattering is due to odderon exchange, predominantly three gluons in a C = $-1$ configuration. At $|t| = 1.0(2.0)$ GeV$^2$ the cross section is predicted to be 7.2(0.2) $\mu$b GeV$^{-2}$. The rate will be about $A\times 100$ s$^{-1}$ where $A$ is the acceptance, at $L = 10^{32}$ cm$^{-2}$s$^{-1}$. If only the forward detectors are read out for events flagged as elastic by our Level 2 trigger the events will be *very* small. Millions of events can be collected with a trigger on the $p$ and $\bar{p}$ with $MM = 0, t_1 = t_2, \Delta\phi = 180°$. The $|t|$-resolution will be $\approx 10^{-2}$ GeV$^2$ and the background should be very small. The elastic scattering events can provide one check on the calibration and $MM, t$ and $\phi$ resolutions of the experiment, although the $x, y$ position of the vertex is not known better than the convoluted transverse beam size ($\sigma_x = \sigma_y \approx 25\mu$m). The longitudinal interaction position $z_o$ is known from the FTC. Rather than taking the full rate (or prescaling) we will probably record only events that appear to be empty (BSC, Miniplugs, TOF all in veto). Then any additional FST tracks will be background (e.g. beam halo) and we will use them to build a "beam halo library" bunch-by-bunch.

### B. Low Mass Exclusive Central Production

Exclusive $p\bar{p} \to p + X + \bar{p}$ where $X$ is a *low mass* state near rapidity $y = 0$ and $+$ represents a rapidity gap $\Delta y >\approx 5.5$ can be studied. As for elastic scattering, there will only be acceptance for large $|t|$. It will be interesting to see whether the mass spectra and flavor composition are the same as when $|t|$ is small (as can be measured already without forward detectors from $+X+$ events). See Appendix III for some physics of special relevance to this proposal which can be done before these pots are installed. At the much lower $\sqrt{s}$ of the SPS (fixed target) the central mass spectra vary rapidly with $\Delta\phi$ [52]. It is interesting to see whether these effects remain at the Tevatron. While we can study central spectra between two large rapidity gaps before we have the VFTD, as we will not then detect the $p$ and $\bar{p}$ we integrate over all $\Delta\phi$. Thus investigations of $\Delta\phi$ dependence will have to wait for the VFTD. One might expect any effect to increase with $|t|$, and we can study that.

### *1. Glueballs and Hybrids*

The central state is a good place to look for glueballs $G$ (produced exclusively, $p\bar{p} \rightarrow p + G + \bar{p}$) and hybrids e.g. $c\bar{c}g$, $b\bar{b}g$. Some papers can be found in the proceedings of the workshop on QCD and Weak Boson Physics in Run II [53].

To search for hybrid states $Q\bar{Q}g$ we will reconstruct effective masses of combinations like $\Upsilon\pi^+\pi^-, \Upsilon\phi$. Note that these events, unlike elastic scattering, have a well-measured central vertex, and often a well measured central mass $M_X$, and therefore provide an excellent calibration of the missing mass scale and resolution.

### *2. $\chi_Q$ states*

States such as $\chi_c^\circ$ (3415 MeV, $\Gamma \approx 10$ MeV) and $\chi_b^\circ$ (9860 MeV, width unknown) have the quantum numbers $I^G J^{PC} = 0^+0^{++}$ (like the vacuum) and hence can be produced in $D\mathbb{P}E$. Little is known about these states apart from their masses from their production in radiative $\psi(2S)$ and $\Upsilon(2S)$ decays. Most of their decay modes are unknown. In particular for $\chi_b^\circ$ the 2000 PDG only gives $\gamma\Upsilon(1S) < 6\%$ with the other 94% unknown. We can trigger on $MM = 9860 \pm 300$ MeV and study the mass spectra of selected likely final states for signs of the $\chi_b^\circ$.

## C. Inclusive $D\mathbb{P}E$

### *1. Spatial extent of color singlets*

By this we mean the transverse spatial distribution of the $\geq 2$ gluons that form the color singlets that are removed from the $p$ and $\bar{p}$ in pomeron exchange. We can not only measure this but determine whether it shrinks as $|t|$ increases, as one might expect. (It is supposed that in large-$|t|$ elastic $pp$ scattering the 3 valence quarks have fluctuated into an unusually close-together configuration.) There are three possible ways (that we are aware of) for measuring the size of the color singlets in $D\mathbb{P}E$.

(a) Measure the total cross section $\sigma_{\mathbb{P}\mathbb{P}}(M_X, t_1, t_2)$ where $M_X$ is the c.m. energy of the $\mathbb{P}\mathbb{P}$ interaction. Because we do not know from first principles the "flux" of the colliding pomeron "beams", this cannot be done without some assumptions. The simplest is to assume soft factorization, in the sense that the $pp\mathbb{P}$ coupling $g_{pp\mathbb{P}}(t)$ is the same in elastic and inelastic processes, at least for the $|t|$-region of our measurements. This may well be true even though we have shown that factorization breaks down for hard diffractive processes.

(b) Double parton scattering, $DPS$. These are two $2 \to 2$ parton scatters in the same interaction, producing (in LO) four jets. The other 4-jet process is double bremsstrahlung, $DBS$. They can be distinguished statistically by the pairwise balance of the $E_T$-vectors in the $DPS$ case. The cross section for DPS depends on the size of the interacting objects (through $\sigma_{DPS} \equiv \frac{\sigma_A \sigma_B}{\sigma_{eff}}$ where $\sigma_{A(B)}$ is the cross section for $2 \to 2$ processes $A$ and $B$, and $\sigma_{eff}$ is an effective area of the overlap of the interacting objects).

(c) Bose-Einstein correlations. Studying correlations at small effective masses between identical bosons, like $\pi^+\pi^+, K^+K^+$ [54] or $K_S^\circ K_S^\circ$ can tell us about the dimension of the region at which the hadrons emerge.

It is our intention to make a second level trigger (see section VIII) on missing mass $MM$ together with other conditions. Thus we could for example select $8 < MM < 11$ GeV for a study of $b\bar{b}g$ hybrids and $\chi_b^\circ$, in parallel with other $MM$ regions for other physics.

### D. Gluon Jet Factory

It has been pointed out by Khoze, Martin and Ryskin [55] that dijets produced by $D\mathbb{P}E$ are almost entirely gluon jets. We quote: "...for the exclusive process the initial $gg$ state obeys special selection rules. Besides being a colour-singlet, for forward outgoing protons the projection of the total angular momentum is $J_z = 0$ along the beam axis. This follows from $P-$ and $T-$invariance and fermion helicity conservation. ... Thus, if we were to neglect the $b-$quark mass, then at leading order we would have no QCD $b\bar{b}$-dijet background (to H) at all." Even without $b$-jet identification the ratio $gg : b\bar{b}$ dijets is expected to be about 3000. Identifying the $b$-jets in the SVX can increase this ratio by an order of magnitude. The light $q\bar{q}$ jets are negligible [56] as long as we suppress large angle gluon radiation by requiring exactly 2 jets. We have measured the cross section for $D\mathbb{P}E$ production of dijets with jet $E_T > 10$ GeV to be a few nb. That corresponds to a million events (times the acceptance) in 1 fb$^{-1}$. In contrast, at present the highest purity $g$-jet sample is 439 jets from $Z \rightarrow b\bar{b}g$ in 5 years of running at LEP1 [57].

## XIII. APPENDIX III: PRE-VFTD STUDIES

There are relevant studies that will be done already before the VFTD is installed [7]. CDF will have one dipole spectrometer arm, new calorimetry (Miniplug) to $|\eta| \approx 5.5$ and "Beam Shower Counters" (BSC) covering approximately $5.5 < |\eta| < 7.5$.

1) Measure the $b\bar{b}$ dijet mass spectrum, $M_{b\bar{b}}$, over the mass range up to 150 GeV to complement the earlier CDF measurement [24]. Using the existing pot spectrometer and the rapidity gap technique we can measure what fraction of these dijets are from single diffraction and what fraction are from $DI\!PE$, as a function of $M_{b\bar{b}}$. What is the angular distribution of the $b$-jets in the $DI\!PE$ case? This studies the background in the $H \to b\bar{b}$ search; but note that the fraction of Higgs bosons that are produced exclusively may be higher than the fraction of all QCD $b\bar{b}$ dijets. Is the ratio $\frac{b-jet}{jet} \approx 3.10^{-4}$ in $DI\!PE$ as predicted?

2) Measure the $l^+l^-$ ($l = e, \mu, \tau$) mass spectrum in the region of $M_{l^+l^-}$ 20-180 GeV, with $\not{E}_T$, studying the associated charged multiplicity $n_{ass}$ on the primary $l^+l^-$ vertex for different mass ranges. How do the results compare with Monte Carlo full event generators of Drell-Yan, Z, $W^+W^-$ and generic (non-$DI\!PE$) Higgs production with leptonic decay? The exclusive $DI\!PE$ Higgs events have $n_{ass} = 0$, and one may observe an excess of events in that bin (or an excess at low $n_{ass}$), which would be *evidence for $DI\!PE$ production of a Higgs.* The only other process we are aware of which could give such events is the two-photon exchange process, but (a) the cross section is much lower (b) this could not produce dilepton events with different flavor, as $H \to W^+W^-$ could (c) the $H \to W^+W^-$ events have a large $\not{E}_T$.

4) Measure the $\gamma\gamma$ mass spectrum, inclusively and when there is one forward rapidity gap, two forward rapidity gaps, and when the $\gamma\gamma$ is exclusive. This will be a normalizer for exclusive and nearly-exclusive Higgs production.